\documentclass[preprint]{aastex}
\usepackage[top=2.5cm, bottom=2.5cm, left=2.5cm, right=2.5cm]{geometry} 
\usepackage{indentfirst} 
\usepackage{amsmath}
\usepackage{natbib}

\shorttitle{Why Do Compact AGNs at High Redshift Scintillate Less?}
\shortauthors{Koay et al.}

\title{Why Do Compact Active Galactic Nuclei at High Redshift Scintillate Less?}

\author{J. Y. Koay\altaffilmark{1*}, J.-P. Macquart\altaffilmark{1}, B. J. Rickett\altaffilmark{2}, H. E. Bignall\altaffilmark{1}, D. L. Jauncey\altaffilmark{3,4}, T. Pursimo\altaffilmark{5}, C. Reynolds\altaffilmark{1}, J. E. J. Lovell\altaffilmark{6}, L. Kedziora-Chudczer\altaffilmark{7}, and R. Ojha\altaffilmark{8,9}}

\altaffiltext{*}{e-mail: kevin.koay@icrar.org}

\altaffiltext{1}{International Centre for Radio Astronomy Research, Curtin University, Bentley, WA 6102, Australia}

\altaffiltext{2}{Department of Electrical and Computer Engineering, University of California, San Diego, La Jolla, CA 92093, USA}

\altaffiltext{3}{CSIRO Astronomy and Space Science, Australia Telescope National Facility, Epping, NSW 1710, Australia}

\altaffiltext{4}{Mount Stromlo Observatory, Weston, ACT 2611, Australia}

\altaffiltext{5}{Nordic Optical Telescope, Apartado 474, 38700 Santa Cruz de La Palma, Spain}

\altaffiltext{6}{School of Mathematics and Physics, University of Tasmania, TAS 7001, Australia}

\altaffiltext{7}{School of Physics and Astrophysics, University of New South Wales, Sydney, NSW 2052, Australia}

\altaffiltext{8}{NASA, Goddard Space Flight Center, Greenbelt, MD 20771, USA}

\altaffiltext{9}{Institute for Astrophysics \& Computational Sciences, The Catholic University of America, 620
Michigan Ave., N.E., Washington, DC 20064, USA}

\clearpage

   \begin{abstract}
      The fraction of compact active galactic nuclei (AGNs) that exhibit interstellar scintillation (ISS) at radio wavelengths, as well as their scintillation amplitudes, have been found to decrease significantly for sources at redshifts $z\gtrsim2$. This can be attributed to an increase in the angular sizes of the $\mu$as-scale cores or a decrease in the flux densities of the compact $\mu$as cores relative to that of the mas-scale components with increasing redshift, possibly arising from (1) the space-time curvature of an expanding Universe, (2) AGN evolution, (3) source selection biases, (4) scatter broadening in the ionized intergalactic medium (IGM) and intervening galaxies, or (5) gravitational lensing. We examine the frequency scaling of this redshift dependence of ISS to determine its origin, using data from a dual-frequency survey of ISS of 128 sources at $0 \lesssim z \lesssim 4$. We present a novel method of analysis which accounts for selection effects in the source sample. We determine that the redshift dependence of ISS is partially linked to the steepening of source spectral indices ($\alpha_{4.9}^{8.4}$) with redshift, caused either by selection biases or AGN evolution, coupled with weaker ISS in the $\alpha_{4.9}^{8.4} < -0.4$ sources. Selecting only the $-0.4 < \alpha_{4.9}^{8.4} < 0.4$ sources, we find that the redshift dependence of ISS is still significant, but is not significantly steeper than the expected $(1+z)^{0.5}$ scaling of source angular sizes due to cosmological expansion for a brightness temperature and flux-limited sample of sources. We find no significant evidence for scatter broadening in the IGM, ruling it out as the main cause of the redshift dependence of ISS. We obtain an upper limit to IGM scatter broadening of $\lesssim 110 \mu$as at 4.9 GHz with 99\% confidence for all lines of sight, and as low as $\lesssim 8 \mu$as for sight-lines to the most compact, $\sim 10 \mu$as sources.

   \end{abstract}


\begin{document}

   \maketitle
   
   \section{Introduction}\label{introduction}
      Radio waves originating from galactic and extragalactic sources are scattered as they propagate through the ionized interstellar medium (ISM) of our own Galaxy, distorting the wavefronts and generating interference patterns on the plane of the Earth. Relative motion between the Earth and the scattering region causes these interference patterns to drift across the observing telescope, and are detected as flux density variations in the source. This phenomenon, known as interstellar scintillation (ISS), has been observed to modulate the flux densities of very compact radio sources such as pulsars and quasars, and has been used as a powerful probe of the physics of the ISM and of the background sources themselves (see \citet{rickett90} and \citet{lazioetal04} for reviews). 

The presence of ISS is highly sensitive to the angular sizes of the sources. In an extended source, its amplitude is suppressed relative to that of a point source. This is because the scattered wavefronts at the observer's plane smear out when integrated over the emission originating from each element of the extended source, similar to the quenching of atmospheric scintillation at optical wavelengths in planets whose angular sizes are larger than that of stars. Since the interfering waves arrive from regions separated by distances up to an order of $\sim 10^6$ km apart at the scattering cloud, the ISM essentially functions as an interstellar interferometer, allowing ISS to probe source sizes down to $\mu$as scales. This provides a resolution orders of magnitude better than ground-based VLBI. In fact, its counterpart, interplanetary scintillation (IPS), has been used in the past to determine the angular sizes of radio sources at sub-arcsecond scales prior to the development of VLBI \citep{littlehewish66, cohenetal67}. 

In the last few decades, the body of evidence linking the intraday variability (IDV) observed in many compact, flat-spectrum Active Galactic Nuclei (AGNs) at cm wavelengths to the physical process of ISS has grown considerably. Time delays between variability patterns have been observed at two widely separated telescopes \citep{jaunceyetal00, dennett-thorpedebruyn02, bignalletal06}. Annual cycles in the variability timescales have also been observed \citep{rickettetal01, jaunceymacquart01, bignalletal03, dennett-thorpedebruyn03, jaunceyetal03}, caused by changes in the velocity of the scattering region relative to the orbital motion of the Earth. The 4.9 GHz, four-epoch Micro-arcsecond Scintillation-Induced Variability (MASIV) Survey \citep{lovelletal03} found 58\% of the observed 443 flat-spectrum sources to exhibit 2 to 10\% rms flux density variations in at least one epoch, shown to be dominated by ISS through a strong correlation with Galactic latitudes as well as line-of-sight Galactic electron column densities \citep{lovelletal08}. 

One of the most surprising results of the MASIV survey was the discovery of a drop in ISS variability amplitudes and the fraction of scintillating sources at redshifts $z \gtrsim 2$ \citep{lovelletal08}. Follow-up observations to the MASIV Survey at 4.9 and 8.4 GHz provided confirmation of this effect at both frequencies \citep[][ hereafter Paper I]{koayetal11}. Although only a sub-sample consisting of 128 sources from the original survey was observed, the sources were monitored over an 11 day period as opposed to the three or four day epochs in the MASIV Survey. We also note that \citet{ofekfrail11} report no significant correlation between ISS amplitude at 1.4 GHz and source redshift, based only on 9 sources for which redshift data were available, in their investigation of the 1.4 GHz variability of sources in the NRAO VLA Sky Survey (NVSS) and the Faint Images of the Radio Sky at Twenty centimeters (FIRST) catalogs. Interestingly, \citet{richardsetal11} found a 3$\sigma$ significance drop in the variability amplitudes of sources at $z > 1$ in their 15 GHz observations of about 1500 Fermi-candidate blazars over a duration of 2 years. While these longer timescale variations at 15 GHz are mainly intrinsic to the sources, it is conceivable that if these high-redshift sources are less compact (in angular and linear scales) than their low-redshift counterparts, they will exhibit lower levels of both ISS and intrinsic variability.    

Determining the origin of this redshift dependence of AGN ISS has important cosmological consequences, potentially allowing the ISS of AGNs to be used as a cosmological probe with the highest angular resolution possible. To understand this redshift dependence, we envision a simple model in which the flat-spectrum AGN consists of an ultra-compact, scintillating core component (estimated to be $\lesssim 150 \mu$as from the MASIV observations) surrounded by more extended, milliarcsecond jet components that do not scintillate. The suppression of ISS at redshifts $z \gtrsim 2$ can therefore be attributed either to an increase in the apparent angular diameters of the core components, or a decrease in the compact fraction of the sources, i.e. the emission from the ultra-compact scintillating component of the source becomes less dominant relative to that of the extended non-scintillating components. Possible interpretations include:

1. \textit{Decrease in observed brightness temperature due to cosmological expansion} \citep{rickettetal07} --- In a sample of brightness temperature-limited sources in the emission frame, as is often assumed for radio-loud AGNs \citep{readhead94}, the mean observed brightness temperature decreases with redshift as a result of the expansion of the Universe. In a flux-limited sample, this results in an increase in the apparent angular diameters of sources with redshift. If this effect dominates, this provides an angular size-redshift relation for extragalactic radio sources that has long been sought after \citep{gurvitsetal99}.  

2. \textit{Evolution of AGN morphologies} --- The drop in ISS can also be explained by a prevalence of sources with lower Doppler boosting factors at high redshift, which would result in either lower source compact fractions or larger angular diameters of the core components.  However, little is currently known about the evolution of AGN core-jet morphologies with redshift, critical in studies of feedback processes in galaxy formation. It is therefore a target of very high resolution observations such as RadioAstron's early science programs. 

3. \textit{Source selection effects} --- It is well known that a flux-limited sample of flat-spectrum AGNs is severely biased with redshift due to source orientation and relativistic beaming \citep{listermarscher97}. Variations in the distribution of intrinsic source luminosities and jet Doppler boosting factors with redshift can lead to a redshift dependence of the apparent mean angular sizes or source compact fractions. Furthermore, a survey at a fixed frequency observes the sources at increasing rest frame emission frequencies with increasing redshift, and thereby observes the optically thick sources at increasing optical depths with increasing redshift, as well as different portions of the spectrum of emission.

4. \textit{Scatter broadening in the ionized intergalactic medium (IGM) and the ISM of intervening galaxies} \citep{rickettetal07} --- Cosmological models demonstrate that supernovae-driven galactic outflows can inject turbulence into the IGM \citep{evoliferrara11}. Such turbulence in the ionized IGM can cause angular broadening due to multipath propagation of the scattered waves. If indeed the redshift dependence of ISS is a result of angular broadening in the IGM, it would present a first direct detection of scattering in the ionized IGM, and would open up a new method of probing the physics of the IGM where $90\%$ of the baryons in the Universe reside \citep{fukugitapeebles04}. As scattering is sensitive to the ionized components, it will complement Lyman-$\alpha$ studies of the neutral component. It may even provide an alternative means of detecting the Warm-Hot component of the IGM, widely believed to be the `missing baryons' in the local Universe based on cosmological hydrodynamical simulations \citep{cenostriker99,daveetal01,cenostriker06}, but which has so far been difficult to detect \citep{bregman07}. This scatter broadening may even occur in the ionized ISM of intervening galaxies, which will provide information on their turbulent properties, although this effect is unlikely to dominate in the majority of the high redshift MASIV sources. 

5. \textit{Gravitational lensing by foreground sources} --- The combined data from the Cosmic Lens All-Sky Survey (CLASS) and the Jodrell Bank VLA Astrometric Survey (JVAS) revealed only 22 gravitational lens systems out of a sample of 16,503 flat-spectrum sources \citep{myersetal03a,myersetal03b}. They were however, searching mainly for multiply imaged sources with arc-second resolution (with follow-up observations using VLBA and MERLIN for confirmation), thus would not have detected any low-level magnification in the sources caused by weak lensing. However, if weak lensing broadens the source images by 10 to 100 $\mu$as, it could supress the ISS of sources at high redshift. Such an explanation would have implications for the distribution of matter (both dark and baryonic) in the low redshift Universe.
 
Three lines of investigation are actively being pursued by the MASIV collaboration and others to better understand this ISS redshift dependence. The first involves a thorough examination of the optical properties of the MASIV sources to detect possible biases due to the presence of sources drawn from different AGN populations (Pursimo et al., submitted). This includes new observations to obtain more reliable redshift estimates and optical identifications (IDs) for the sources to complement archival data. New spectroscopic redshifts and IDs for many of the MASIV sources in which such data were not available, have also been obtained. 

The second line of investigation makes use of VLBI data to examine the mas-scale structures of the sources to determine their effects on ISS \citep{ojhaetal04a}. While it has been found that scintillating sources are more core dominated than the non-scintillating sources at mas-scales \citep{ojhaetal04b}, how their mas compact fractions scale with redshift is still being investigated. Multi-frequency VLBI observations to study possible angular broadening in a subsample containing 49 MASIV sources \citep{ojhaetal06} found no evidence of scatter broadening in the IGM at the resolution probed \citep{lazioetal08}, providing an upper limit of $500 \mu$as at 1.0 GHz.   

The third key to understanding the redshift dependence of ISS, which is the focus of this present paper, is to examine how this redshift dependence scales with observing frequency. The motivation is that scatter broadening scales roughly with ${\nu}^{-2}$, whereas intrinsic source size effects scale with ${\nu}^{-1}$ for a synchrotron self-absorbed source. On the other hand, the effects of cosmological expansion and gravitational lensing are achromatic. Therefore, while we generally expect the amplitude of weak ISS to decrease when we go to higher observing frequencies (see \cite{narayan92} for a review of the different regimes of ISS), we will observe either a similar or weaker redshift scaling of ISS amplitudes depending on which interpretation is correct. At the very least, it will allow us to rule out some of the interpretations listed above, or place strong constraints on them. 

The dual-frequency MASIV follow-up observations of ISS provide us with the data most suited for such an investigation. Indeed, the redshift dependence was found to be marginally weaker at 8.4 GHz as compared to 4.9 GHz, which can be interpreted as due to weaker IGM scatter broadening of the $z \gtrsim 2$ sources at the higher frequency (see Paper I). However, the presence of subtle selection effects such as those mentioned above, along with the complexity of the ionized ISM and the sources themselves, complicates the interpretation. 

In this follow-up paper, we present a comprehensive analysis of the data, taking these selection effects into consideration. We also make use of more accurate models and Monte-Carlo simulations to interpret the data. We provide a brief summary of the observations and characterization of source variability in Section~\ref{methodology}. Section~\ref{analysis} then delves into the analysis and interpretation of the results. Our conclusions are summarized in Section~\ref{conclusion}.

   \section{Observations and Source Variability Characterization}\label{methodology} 
      We observed a sub-sample of 140 sources drawn from the original MASIV Survey over a duration of 11 days using the VLA. 70 of these sources have redshifts of $z < 2$ (we refer to them as the low-redshift sources) while another 70 have redshifts of $z > 2$ (the high-redshift sources). The telescope was divided into 2 subarrays, one observing at 4.9 GHz and the other observing at 8.4 GHz simultaneously. Each source was observed for 1 minute at $\approx 2$ hour intervals. The visibilities were then coherently averaged over each 1 minute scan to produce the time-series data (lightcurves) for each source at both frequencies. 

We used the structure function, $D_{obs}(\tau)$, as a standard characterization of the source variability amplitudes at various timelags $\tau$, given by:
\begin{equation}\label{sf}
D_{obs}(\tau)=\left< \left[S(t+\tau) - S(t) \right]^{2} \right> ,
\end{equation}
where $S(t)$ is the flux density of the source at time $t$, normalized by its mean flux density calculated from the entire 11-day period. The angular brackets indicate averaging over time, $t$. We then fit the following model to $D_{obs}(\tau)$ for each source:
\begin{equation}\label{sfmodel}
D_{mod}(\tau)=D_{sat}\dfrac{\tau}{\tau+{\tau}_{char}}+D_{noise},
\end{equation}
assuming that ISS approaches a stationary stochastic process when observed over a duration much longer than its characteristic timescale. $D_{sat}$ is the value at which $D_{mod}(\tau)$ saturates, and $\tau_{char}$ is the characteristic timescale at which the structure function (SF) reaches half of its value at saturation. Any variability caused by instrumental and systematic errors were assumed to contribute a white additive noise, $D_{noise}$, across all time-lags. We estimated $D_{noise}$ as a quadratic sum of the flux independent errors i.e. system noise and confusion, and the flux dependent calibration errors, which we then subtracted from $D_{mod}(\tau)$ across all time-lags to obtain the `true' variability, $D(\tau)$. 12 sources were removed from our sample due to large errors that were not well quantified by our estimation of $D_{noise}$. We then used $D(\tau)$ at a time-lag of 4 days, $D(\rm 4d)$, for all subsequent analyses as a standard characterization of source variability amplitudes. In this paper, we denote $D(\rm 4d)$ at 4.9 GHz and 8.4 GHz as $D_{4.9}(\rm 4d)$ and $D_{8.4}(\rm 4d)$ respectively. Significant correlation between $D(\tau)$ and line-of-sight H$\alpha$ intensities at timelags of 1 to 7 days (see Paper I) confirm that ISS dominates $D(\tau)$ at these timescales. 

We refer the interested reader to Paper I for an extensive elucidation of the techniques and analyses summarized here. The variability amplitudes of the 128 sources used in our sample are listed in Table~\ref{data} along with other source properties used in our analyses.

   \section{Analysis and Interpretation}\label{analysis}  
      
      \subsection{Selection Effects}\label{selection}
         The results of the MASIV Survey and the follow-up observations show that the presence of ISS is strongly correlated with source spectral indices, mean flux densities and line-of-sight H$\alpha$ intensities \citep{lovelletal08}. We therefore examine here whether these parameters are similarly distributed across the low and high-redshift samples, as well as other selection effects that may bias the interpretation of the results. 

The sources in the original MASIV sample were selected to be flat-spectrum, core-dominated AGNs, based on the spectral index criterion of $\alpha_{1.4}^{8.4} > -0.3$ (where $S \propto \nu^{\alpha}$). The cutoff was set at a higher than usual value of $-0.3$ (where $-0.4$ or $-0.5$ is normally used), to avoid the tail end of the distribution of the classical steep-spectrum sources peaked at $\alpha \sim -0.7$ \citep{scheuerwilliams68}, considering that the $\alpha_{1.4}^{8.4}$ values were estimated from non-coeval mean flux densities at different frequencies. However, we found 30 sources with spectral indices of $\alpha_{4.9}^{8.4} < -0.3$ in the present sample, of which 15 have $\alpha_{4.9}^{8.4} < -0.4$, calculated from the coeval mean flux densities in our 11 day observations. 

\begin{figure}[!htp]
     \begin{center}
     \includegraphics[scale=0.7]{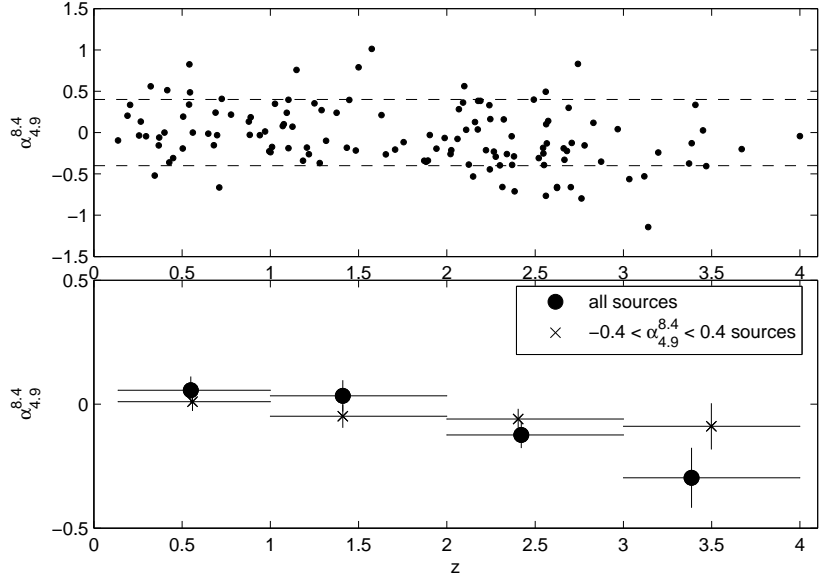}
     \end{center}
     \caption{{The top panel shows a scatter plot of source spectral indices against source redshift for all 128 sources. The horizontal dashed lines indicate $\alpha^{4.9}_{8.4} = -0.4$ and $\alpha^{4.9}_{8.4} = 0.4$. The bottom panel shows the mean spectral indices in four redshift bins for all 128 sources as well as for the $-0.4 < \alpha_{4.9}^{8.4} < 0.4$ sources. The error bars in the binned plots indicate 1$\sigma$ errors in the mean.} \label{spec_vs_z_before}}
     \end{figure}

We also found that the redshift dependence of ISS in our sample can be at least partially linked to the steepening of the mean values of $\alpha_{4.9}^{8.4}$ with redshift (see Figure~\ref{spec_vs_z_before}). The Kolmogorov-Smirnov (K-S) test rejects the null hypothesis, that $\alpha_{4.9}^{8.4}$ of the low and high-redshift samples are drawn from the same parent population, with a $7.4 \times 10^{-3}$ probability that this result was obtained by chance (here and in all subsequent analyses, we consider probabilities below the standard value of 0.05 to be statistically significant, while probabilities above this value are considered not significant). Of the 15 $\alpha_{4.9}^{8.4} < -0.4$ sources, which are known to scintillate less than the $\alpha_{4.9}^{8.4} > -0.4$ sources (Paper I), 13 are found at $z > 2$. Furthermore, eight of the 11 $\alpha_{4.9}^{8.4} > 0.4$ sources are found at $z < 2$, and there are indications that their scintillation amplitudes may be larger than that of the $-0.4 < \alpha_{4.9}^{8.4} < 0.4$ sources (Paper I). This $z$-$\alpha_{4.9}^{8.4}$ correlation in itself is of great interest and we defer a full discussion of its significance to Section~\ref{z-specindex}. 

Since $\alpha_{4.9}^{8.4}$ is based on the coeval flux densities of the follow-up observations, we have used $-0.4$ as a lower cutoff for the selection of sources. This allows us to retain slightly more sources for better statistical representation. Also, an examination of the 15 sources with $-0.4 < \alpha_{4.9}^{8.4} < -0.3$ finds that they are distributed roughly equally, with 8 at $z < 2$ and 7 at $z > 2$. We therefore remove only 11 $\alpha^{4.9}_{8.4} > 0.4$ and 15 $\alpha_{4.9}^{8.4} < -0.4$ (a total of 26) sources from our sample, after which the K-S test shows that the distribution of $\alpha_{4.9}^{8.4}$ in the high and low-redshift samples no longer differ significantly. In any case, we note that all subsequent analyses in this and the following sections were performed on both the $-0.4 < \alpha_{4.9}^{8.4} < 0.4$ and $-0.3 < \alpha_{4.9}^{8.4} < 0.4$ samples, for which we found no differences in the conclusions. From here onwards, we present only the results for the $-0.4 < \alpha_{4.9}^{8.4} < 0.4$ sample of 102 sources, comprising 46 sources at high redshift and 56 sources at low redshift.   

\begin{figure}[!htp]
     \begin{center}
     \includegraphics[scale=0.8]{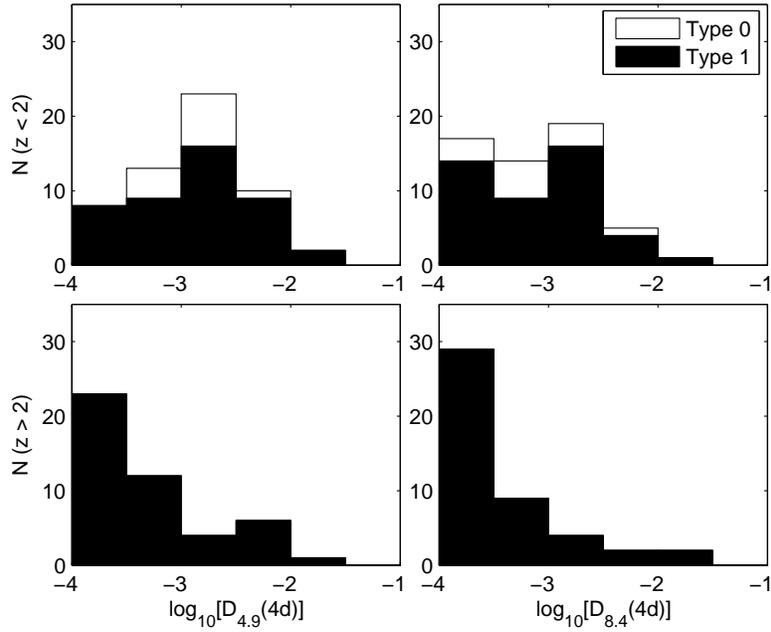}
     \end{center}
     \caption{{Distribution of $D_{4.9}({\rm 4d})$ (left) and $D_{8.4}({\rm 4d})$ (right) in the low (top) and high (bottom) redshift sample of sources. Only the 102 $-0.4 < \alpha_{4.9}^{8.4} < 0.4$ sources are shown, classified into Type 0 or Type 1 AGNs based on their optical IDs. Since $D_{noise} \approx 1 \times 10^{-4}$ on average at both frequencies, we include all sources with $D({\rm 4d}) < 1 \times 10^{-4}$ in the $-4 < {\rm log_{10}}[D({\rm 4d})] < -3.5$ bin. The high redshift sources are significantly less variable than their low redshift counterparts.} \label{histoD4d}}
     \end{figure}

Considering only the $-0.4 < \alpha_{4.9}^{8.4} < 0.4$ sources, we found that the drop in ISS amplitudes at high redshift remains statistically significant. The K-S test confirms that the variability amplitudes of the high-redshift sources, characterized by $D_{4.9}(\rm 4d)$ and $D_{8.4}(\rm 4d)$, are significantly smaller than that of the low-redshift sources, with a probability of $5.1 \times 10^{-5}$ that this occured by chance. This is clearly seen in the histograms of Figure~\ref{histoD4d}. 

We note that in the selection of sources for the MASIV follow-up observations, the $z < 2$ sample was biased towards the variable sources, while all sources with known redshifts were selected for the $z > 2$ sample. Although one could argue that the fraction of scintillating sources at low redshift is higher than that at high redshift anyway, as found in the MASIV survey \citep{lovelletal08}, this introduces another possible selection effect into the source sample. However, the significant decrease in ISS amplitudes at high redshift of the $-0.4 < \alpha_{4.9}^{8.4} < 0.4$ sources cannot be attributed solely to this selection effect, as the significance is retained even when only the most variable sources are selected. K-S tests show that $D_{4.9}({\rm 4d})$ and $D_{8.4}({\rm 4d})$ in the $z < 2$ sources are still significantly larger than their counterparts in the $z > 2$ sources when considering only the 72 sources where $D({\rm 4d}) \geq 2 \times D_{noise}$ (equivalent to $\geq 3\sigma$ variability amplitudes) at both frequencies, with probabilities of $1.3 \times 10^{-2}$ and $2.2 \times 10^{-2}$ that they occured by chance. These 72 sources were drawn from only the $-0.4 < \alpha_{4.9}^{8.4} < 0.4$ sample. 

The sources in our sample were also carefully selected so that their mean flux densities and line-of-sight H$\alpha$ intensities would be evenly distributed in the high and low-redshift samples, but a large fraction of these sources are variable, so their mean flux densities may have changed. As a check, we performed the K-S test which found no statistically significant differences in the distribution of mean flux densities and line-of-sight H$\alpha$ intensities in the low and high-redshift samples. This is true before and after the removal of the 26 $\alpha_{4.9}^{8.4} > 0.4$ and $\alpha_{4.9}^{8.4} < -0.4$ sources, as well as for the 72 sources with $\geq 3\sigma$ variability.

Another possible source of selection bias is the presence of sources with different optical IDs in the sample. 45 out of the 46 high-redshift sources are identified as flat-spectrum radio-loud quasars (FSRQs), while the other source has yet to be identified, most likely an FSRQ. In the low-redshift sample however, 42 sources are identified as FSRQs, 12 are BL Lac objects, and 2 are Seyfert 1 galaxies. In standard AGN unification schemes \citep{urrypadovani95}, FSRQs and Seyfert 1 galaxies are classified as Type 1 AGNs, with both having broad emission lines. They differ only in their radio and optical luminosities, the significance of which is historical rather than physical, so we group them together in our analysis. On the other hand, BL Lacs are classified as Type 0 AGNs due to their weak or lack of emission lines. BL Lacs have been observed to be more variable than FSRQs, intrinsically \citep{richardsetal11} as well as in terms of their ISS (Pursimo et al., submitted). We therefore carry out our analysis with and without the inclusion of the BL Lacs to determine if they affect the interpretation of the data.

While biases caused by the aforementioned parameters can be mitigated through the careful selection of sources, there are other selection effects that are unavoidable in a survey such as this. These selection effects can increase or decrease ISS with redshift, biasing the result either way. For example, the sources are observed at increasing rest-frame emission frequency with increasing redshift. For an optically thick synchrotron self-absorbed source, the source size decreases with increasing rest-frame frequency.

\begin{figure}[!htp]
     \begin{center}
     \includegraphics[scale=0.8]{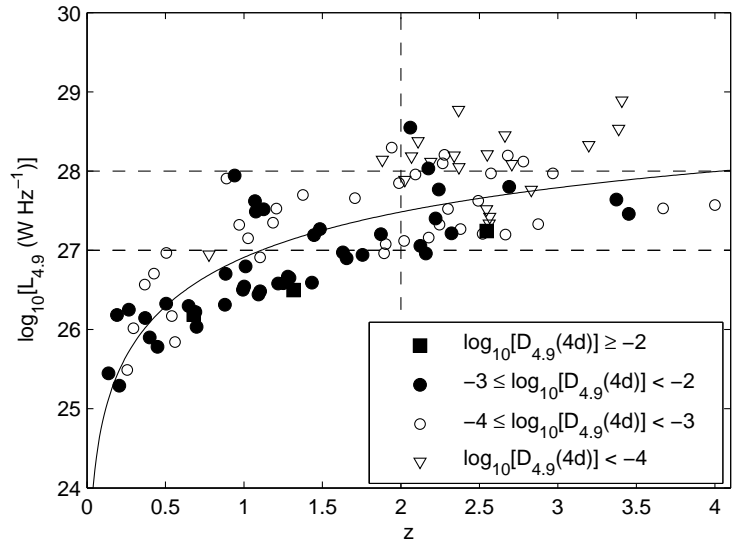}
     \end{center}
     \caption{{4.9 GHz spectral luminosities plotted against source redshifts, with sources separated into four $D_{4.9}({\rm 4d})$} bins. Only the 102 $-0.4 < \alpha_{4.9}^{8.4} < 0.4$ sources are included. The solid curve gives the $L_{4.9}$-$z$ relation for a 0.3 Jy source, separating the strong ($S_{4.9} \geq 0.3$ Jy) sample of sources from the weak ($S_{4.9} < 0.3$ Jy) sample of sources. The vertical dashed line indicates the redshift cutoff of $D_{4.9}({\rm 4d})$ at $z \sim 2$, while the dashed horizontal lines indicate possible luminosity cutoffs of $D_{4.9}({\rm 4d})$ at $L_{4.9} \sim 10^{28}\, {\rm WHz^{-1}}$  and $L_{4.9} \sim 10^{27}\, {\rm WHz^{-1}}$ for the strong and weak sources respectively. \label{L_vs_z}}
     \end{figure}

Furthermore, a flux-limited survey will always be affected by the Malmquist bias arising 
from the scaling of source spectral luminosity, $L_{\nu}$, with redshift. In fact, this perceived redshift dependence of ISS may even be interpreted as a luminosity dependence. It is possible that the higher-luminosity sources are intrinsically larger, and may therefore scintillate less. The plot of the 4.9 GHz spectral luminosities (calculated using $H_{0} = 70{\,\rm kms^{-1} Mpc^{-1}}$, $\Omega_{M} = 0.27$, $\Omega_{\Lambda} = 0.73$ and assuming isotropic emission) against source redshifts (Figure~\ref{L_vs_z}) shows that at any particular redshift, there is a luminosity dependence of ISS because the stronger sources ($S_{4.9} \geq 0.3$ Jy) scintillate less than the weaker sources ($S_{4.9} < 0.3$ Jy). There is a visible redshift cutoff of $D_{4.9}(\rm 4d)$ at $z \gtrsim 2$ for both the weak and strong sources. There are also possible luminosity cutoffs at $L_{4.9} \sim 10^{28} {\rm WHz^{-1}}$ and $L_{4.9} \sim 10^{27} {\rm WHz^{-1}}$ for the strong and weak sources respectively. However, it is difficult in reality to determine if the ISS amplitudes of AGNs (in our present sample at least) have a redshift cutoff, luminosity cutoff or both, since $S_{\nu}$, $z$ and $L_{\nu}$ are inter-dependent. This problem is further compounded for relativistically beamed sources, where Doppler boosting complicates the definition of the intrinsic source luminosities. Intrinsically more luminous sources can be detected at larger angles of orientation to the line of sight, and may thus have lower Doppler boosting factors which decrease their compact fractions with redshift. On the other hand, if the high-redshift sources are dominated by highly Doppler-boosted sources with intrinsic luminosities comparable to their low-redshift counterparts, the compact fractions of the high-redshift sources may be larger.

      \subsection{Structure Function Ratios}\label{ratios}
         We present a method to alleviate the effects of the selection biases discussed in Section~\ref{selection}, in particular the redshift dependence of source luminosities, Doppler boosting factors and rest-frame emission frequencies. Instead of comparing the redshift dependences of the mean values of $D_{4.9}(\rm 4d)$ and $D_{8.4}(\rm 4d)$ separately, as was done in Paper I, it is more instructive to obtain the ratio of $D_{8.4}(\rm 4d)$ to $D_{4.9}(\rm 4d)$ for each individual source, then compare the mean and distribution of the ratios at high and low redshift. This normalizes each source by itself, thereby reducing the dependence of the results on these source parameters if they are frequency independent or have known frequency dependences. The rest-frame emission frequencies also become irrelevant, since we are only interested in the ratio of the two observing frequencies. Parameters involving the properties of the ISM are also factored out, again assuming all these source parameters to be frequency independent. We provide a mathematical justification for these claims in Section~\ref{theory1} and present our results in Section~\ref{results1}.

\subsubsection{Theoretical Basis}\label{theory1}

To obtain theoretical estimates of the structure function ratios for comparison with the observational data, we make use of standard ISS models in which the scattering region is approximated as a thin, phase-changing screen with an isotropic Kolmogorov spectrum. The quantity of interest is the spatial coherence, $\Gamma_{4}(r;\nu)$, of the flux measured at two locations separated by a distance $r$ on the Earth at a frequency of $\nu$ (\citet{goodmannarayan89} provide the detailed formalisms). The model assumes that the phase structure function at the scattering screen is isotropic at the length-scales of interest and does not vary at the timescales ($\tau$) of interest, so that the spatial coherence can be simply related to the temporal coherence as measured by a single telescope by equating $r = v_{s}\tau$, where $v_{s}$ is the transverse velocity of the scattering screen relative to the Earth. We then compute $D({\rm 4d}) = 2[\Gamma_{4}(0;\nu) - \Gamma_{4}(r = v_{s}\cdot {\rm 4d};\nu)]$. 

Analytical solutions for $\Gamma_{4}(r;\nu)$ in the very weak and very strong scintillation regimes are given in \citet{colesetal87} and \citet{narayan92}. They use the modulation index, $m$, defined as the rms variations as a percentage of the mean flux density of the source, to quantify the variability amplitude of the source. Assuming that the structure functions saturate within 4 days, $D({\rm 4d}) \approx 2m^{2}$. In weak scintillation ($\nu \gg \nu_{t}$, where $\nu_{t}$ is the transition frequency between weak and strong ISS), the modulation index of a point source is given in the following form by \citet{walker98}:
\begin{equation}\label{mpw}
m_{p,w} = \left( \dfrac{\nu_{t}}{\nu}\right)^{17/12},
\end{equation}  
For strong refractive scintillation ($\nu \ll \nu_{t}$), the point source modulation index is \citep{walker98}:
\begin{equation}\label{mpr}
m_{p,r} = \left( \dfrac{\nu}{\nu_{t}}\right)^{17/30}.
\end{equation}
The modulation index of an extended source is then further suppressed relative to that of a point source by a factor that is dependent on the apparent angular size of the source, $\theta$, as it appears to the scattering screen \citep{walker98}:
\begin{equation}\label{m}
m = m_{p}\left( \dfrac{\theta_{ext}}{\theta}\right)^{7/6},
\end{equation}
where $m_{p}$ is equivalent to $m_{p,w}$ or $m_{p,r}$. $\theta_{ext}$ is the angular size above which the source can no longer be considered a point source. For weak ISS, $\theta_{ext}$ is equivalent to the angular size of the first Fresnel zone at the scattering screen, given by $\theta_{F} = \sqrt{c/(2 \pi \nu D_{\rm ISM})} \propto {\nu}^{-0.5}$, where $D_{\rm ISM}$ is the distance from the Earth to the scattering screen and $c$ is the speed of light. For strong refractive ISS, $\theta_{ext}$ is the refractive scale of the density inhomogeneities at the scattering screen, given by $\theta_{r} \propto {\nu}^{-2.2}$. 

The ratio of $D_{8.4}(\rm 4d)$ to $D_{4.9}(\rm 4d)$, which we denote as $R_{D}$, can be calculated using the asymptotic limits of Equations~\ref{mpw} to~\ref{m} to obtain:
\begin{equation}\label{RDweak}
R_D \approx \frac{m_{8.4}^2}{m_{4.9}^2} = 0.217\, {\left( \frac{\theta_{F,8.4}}{\theta_{F,4.9}} \right)}^{7/3} {\left( \frac{\theta_{4.9}}{\theta_{8.4}} \right)}^{7/3},
 \end{equation} 
for an extended source in the weak ISS regime, and: 
 \begin{equation}\label{RDstrong}
R_D \approx 1.842\, {\left( \frac{\theta_{r,8.4}}{\theta_{r,4.9}} \right)}^{7/3} {\left( \frac{\theta_{4.9}}{\theta_{8.4}} \right)}^{7/3},
 \end{equation} 
for an extended source in the strong refractive ISS regime. Any anisotropic properties of the ISM (i.e. the strength of turbulence and distance to the scattering screen), encapsulated in the terms $\nu_t$, $\theta_{F}$ and $\theta_{r}$, either cancel out or have known frequency dependences. $R_{D}$ is therefore sensitive only to the ratio $\theta_{4.9}/\theta_{8.4}$.

The apparent source size, whose frequency scaling is dependent upon whether it is dominated by intrinsic effects or scatter broadening, can be modeled as:
\begin{equation}\label{theta}
 \theta \sim \sqrt{\theta_{src}^{2} + \theta_{scat}^{2}} \propto \nu^{-\beta},
 \end{equation} 
where $\theta_{src}$ is the intrinsic source angular size, and $\theta_{scat}$ represents additional contributions due to scatter broadening in the ISM or the IGM. If intrinsic source size effects dominate, $\theta \sim \theta_{src}$, and any source dependent parameters that $\theta_{src}$ is a function of, such as the luminosity, compact fraction and Doppler boosting factor, cancel out in Equations~\ref{RDweak} and \ref{RDstrong} assuming that they are frequency independent. $R_{D}$ is therefore transparent to any redshift dependences of these parameters. The ratio of the emission-frame frequencies is a constant across all redshifts for a fixed pair of observing frequencies, so $R_{D}$ is also insensitive to source properties that vary with increasing emission-frame frequencies. Typically, $\beta \sim 1$ for a synchrotron self-absorbed source, while $\beta \sim 0$ if the source size is frequency independent. On the other hand, $\beta \sim 2.2$ for a scattering screen with Kolmogorov turbulence, if the source size is dominated by scatter broadening so that $\theta \sim \theta_{scat}$. 

We calculated $R_{D}$ for different values of $\beta$ in the asymptotically weak and strong refractive ISS regimes. In the case of a point source ($\theta < \theta_{ext}$), $R_{D} \sim 0.2$ and $R_{D} \sim 1.8$ in the weak ISS and strong refractive ISS regimes respectively, and is independent of $\beta$. At our observing frequencies and for sight-lines through mid-Galactic latitudes however, $\theta > \theta_{ext}$ for AGN in general. In this case, $R_{D} \lesssim 0.4$ when intrinsic source size effects dominate ($\beta \leq 1$) and $R_{D} \sim 1.8$ when scatter broadening dominates ($\beta = 2.2$), for both weak and strong ISS. 

\begin{figure}[!htp]
     \begin{center}
     \includegraphics[scale=1.0]{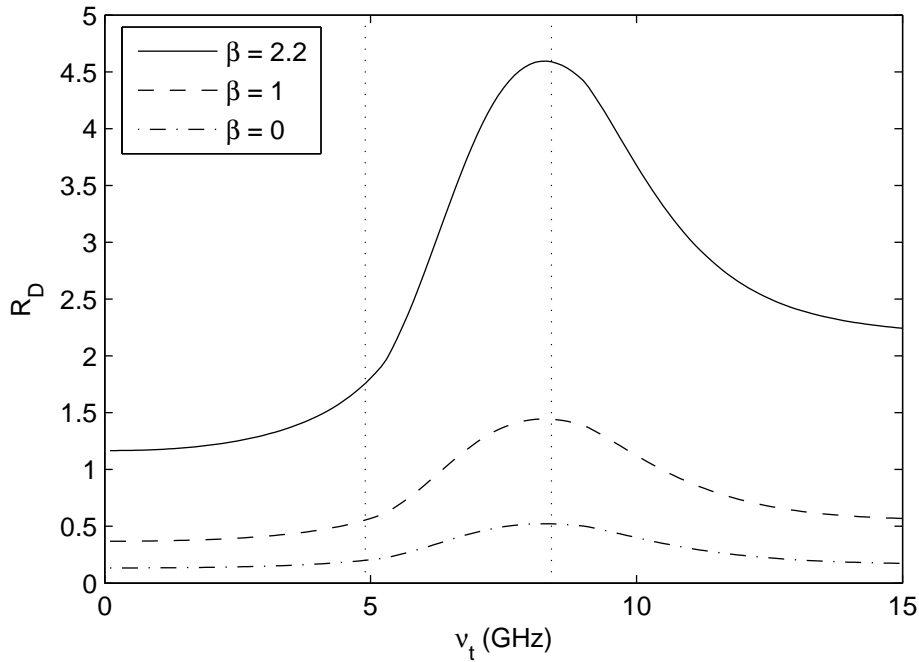}
     \end{center}
     \caption{{The structure function ratio, $R_{D}$, calculated for $\beta = 0, 1$ and $2.2$ using the fitting function in \citet{goodmannarayan06} and plotted against the transition frequency between weak and strong scattering. We have adopted values of $v_{s} = 50\,{\rm kms}^{-1}$, $\theta = 100 \, \mu$as at 4.9 GHz, and a scattering screen distance of $D_{\rm ISM} = 500 {\rm pc}$. The vertical dashed lines represent $\nu = 4.9$ GHz and $\nu = 8.4$ GHz respectively.} \label{ratio_betatransition}}
     \end{figure}

In the intermediate scattering regimes typical of our observations, where there are no analytical solutions, we make use of the fitting functions for $\Gamma_{4}(r;\nu)$ provided by \citet{goodmannarayan06} based on numerical computations, which allow us to calculate $R_{D}$ when $4.9 \,{\rm GHz} \lesssim \nu_{t} \lesssim 8.4 \,{\rm GHz}$. Figure~\ref{ratio_betatransition} demonstrates how $R_{D}$ varies with $\nu_{t}$ for $\beta = 0, 1$ and $2.2$. 

While the analytical approach provides a better understanding of the physics involved in the analysis of the SF ratios, it assumes asymptotically weak or strong refractive ISS of the sources. Furthermore, it assumes that the characteristic timescales of ISS are less than four days so that $D({\rm 4d})$ is well approximated by $2m^2$. The ISS timescales can vary with observing frequency, which affects $R_{D}$ if the SFs have yet to saturate at one or both frequencies. We know from Paper I that $\sim 20\%$ of the sources have ISS timescales of more than four days on at least one frequency. On the other hand, the \citet{goodmannarayan06} fitting function makes no such assumption about the ISS timescales and simply calculates $\Gamma_{4}(v_{s}\cdot {\rm 4d};\nu)$. However, when the ISS timescales are longer than four days, we note that $R_{D}$ estimated from the fitting function becomes sensitive to the various scattering screen and source parameters, which will be true of the observations as well. In interpreting our data, we mainly used the fitting functions for comparisons with the observations.   

\subsubsection{Observational Results}\label{results1}

The histogram of $R_{D}$ as obtained from our observations exhibits a well defined peak in the $0.25 < R_{D} < 0.50$ bin (Figure~\ref{histoSFxSFc}), consistent with the weak ISS of sources dominated by intrinsic source size effects so that $\beta \leq 1$ (see Figure~\ref{ratio_betatransition}). As $R_{D}$ is inaccurate for sources whose $D(\rm 4d)$ is close to or lower than the noise floor at one or both frequencies, the histogram includes only the 72 sources with $\geq 3\sigma$ variability amplitudes at both frequencies, selected from the $-0.4 < \alpha_{4.9}^{8.4} < 0.4$ sample, with 48 sources at low redshift and 24 at high redshift. Of the 48 low redshift sources, 11 are Type 0 AGNs (BL Lacs) and 37 are Type 1 AGNs (FSRQs and Seyfert 1 galaxies).

However, the tail towards larger values of $R_{D}$ indicates the presence of at least another important effect. Three effects potentially increase $R_{D}$ so that it becomes comparable to or greater than unity. One is if $\nu_{t} \gtrsim 5$ GHz, so that the sources are scintillating in the strong ISS regime or at the transition between weak and strong ISS at one or both frequencies. The second possibility is that $\beta > 1$ due to scatter broadening at a second, more distant screen before the waves arrive at the scintillation-inducing screen. This second scattering screen can be Galactic or extragalactic. The third possibility is that the SFs have yet to saturate within 4 days, so that the assumption of $D({\rm 4d}) \approx 2m^{2}$ no longer holds. Since $\tau_{char} \propto \theta$, and assuming $\theta \propto \nu^{-1}$, the scintillation timescales are shorter at 8.4 GHz. This causes the 8.4 GHz SFs to rise faster and saturate earlier in comparison to that at 4.9 GHz, thereby increasing $R_{D}$.

We determined that the sources with $R_{D} \gtrsim 1$ are not significantly scatter broadened, based on an examination of the $R_{D}$ values in the weak ($S_{4.9} < 0.3$ Jy) and strong ($S_{4.9} \geq 0.3$ Jy) sample of sources. We can assume that the weak and strong sources have similar mean intrinsic brightness temperatures, so that a $\sim 0.1$ Jy source tends to have a smaller angular diameter than a $\sim 1.0$ Jy source. This is not an unreasonable assumption, as it explains why the weak sources have been observed to scintillate more than the stronger sources \citep{lovelletal08}. Additionally, the lower ISS amplitudes in the strong sample of sources cannot be attributed to the presence of stronger mas-scale jet components, as confirmed by VLBI observations that found no significant difference in the mas-scale morphologies of the strong and weak flux density sources \citep{ojhaetal04b}. From Equation~\ref{theta}, we see that the effects of scatter broadening will be more dominant in sources with smaller intrinsic angular sizes. Therefore, if the sources in our sample are scatter broadened, we should observe higher values of $R_{D}$ in the weaker sources than in the stronger sources. K-S tests do not show that the weak sources have $R_{D}$ values significantly higher than that of the strong sources. This is true when all 72 sources are considered and when only the Type 1 AGNs are considered. In fact, the converse is true. Figure~\ref{sfxsfcweakstrong} shows scatter plots of $D_{8.4}(\rm 4d)$ against $D_{4.9}(\rm 4d)$ for the 72 sources with $D(\rm 4d) \geq 3\sigma$, classifying the sources into weak and strong samples. The dotted lines have slopes of 0.4 and 1.8, representing the possible values of $R_{D}$ in the asymptotic weak and strong ISS regimes. The solid lines represent linear least-square fits for the two source categories. In obtaining the fits, each source is weighted by a factor:
\begin{equation}\label{weight}
w = \left({\sigma_{D,4.9}^{2}+\sigma_{D,8.4}^{2}} \right)^{-0.5},
\end{equation}
where $\sigma_{D,4.9}$ and $\sigma_{D,8.4}$ are the normalized errors in $D_{4.9}(\rm 4d)$ and $D_{8.4}(\rm 4d)$ respectively. This means that sources that have smaller errors in $D(\rm 4d)$ have larger weights in the fitting process. The dashed lines are the 99\% confidence bounds for those fits. It can be seen that the strong sources have a larger fitted $R_D$ of $1.06 \pm 0.34$ as compared to a fitted $R_D$ of $0.44 \pm 0.18$ for the weak sources at 99\% confidence.  

We found $R_{D}$ to be influenced by the strength of ISS as indicated by the line-of-sight H$\alpha$ intensity to each source, obtained from the Wisconsin H-Alpha Mapper (WHAM) Northern Sky Survey \citep{haffneretal03}. The line-of-sight H$\alpha$ intensity is in units of Rayleighs (R), and we denote it as $I_{\alpha}$. The K-S test confirms that $R_{D}$ in the $I_{\alpha} \geq 5$ R sample is significantly larger than that of the $I_{\alpha} < 5.0$ R sample, with a $1.3 \times 10^{-2}$ probability that this occured by chance when all sources are considered, and a probability of $4.0 \times 10^{-2}$ when only the Type 1 AGNs are considered. The scatter plots and fits of $R_{D}$ in Figure~\ref{sfxsfchalpha} also show that sources with $I_{\alpha} \geq 5$ R tend to have larger values of $R_{D}$. The fitted $R_{D}$ is found to be 0.99 $\pm$ 0.59 for $I_{\alpha} \geq 5$ R as compared to 0.35 $\pm$ 0.08 for $I_{\alpha} < 5.0$ R at 99\% confidence. The sight-lines with larger H$\alpha$ intensities have higher electron column densities and are at lower Galactic latitudes where the sources are seen through thicker regions of the Galaxy. Therefore, we expect the transition frequencies to be higher through these sight-lines. This demonstrates that sources with $R_{D} \gtrsim 1$ are scintillating in the strong ISS regime or at the transition between weak and strong ISS. Additionally, the five most variable sources with $R_{D} > 1$ all have $I_{\alpha} \geq 5$ R, consistent with scintillation amplitudes being the highest at the transition frequency between weak and strong ISS.  

In comparing the values of $R_{D}$ in the $z < 2$ and $z > 2$ subsamples, we rule out scatter broadening in the IGM as the origin of the redshift dependence of ISS, in our present sample at least. Using the K-S test, we find that $R_{D}$ in the high-redshift sample is not significantly higher than $R_{D}$ in the low-redshift sample considering all sources and only the Type 1 AGNs. Figure~\ref{sfxsfcredshift} shows only a marginal increase in the fitted $R_{D}$ of sources in the $z > 2$ sample as compared to the $z < 2$ sample, with 0.83 $\pm$ 0.35 for $z > 2$ and 0.41 $\pm$ 0.16 for $z < 2$ (99\% confidence intervals). The fitted $R_{D}$ values at low and high redshift are calculated separately for the weak and strong sample of sources, summarized in Table~\ref{variousRDz}. The redshift dependence of $R_{D}$ in the weak sources is not significantly steeper than that of the strong sources, providing more evidence against scatter broadening in the IGM or that it has any redshift dependence at the resolution of our observations. 

In fact, the marginal increase in $R_{D}$ with redshift is most plausibly attributed to intrinsic source size effects, since the stronger sources that have larger intrinsic angular sizes also have larger fitted $R_{D}$. We explain this increase in $R_{D}$ with redshift in Section~\ref{expansion}, and discuss the implications of these results on IGM scattering properties in Section~\ref{constraints}.

      \subsection{Decrease in Observed Brightness Temperature Due to Cosmological Expansion}\label{expansion}
         We propose that the suppression of ISS at $z \gtrsim 2$, considering only the $-0.4 < \alpha_{4.9}^{8.4} < 0.4$ sources, can be sufficiently explained by the decrease in observed brightness temperature of a flux-limited sample of sources due to cosmological expansion. The angular size of a source, $\theta_{src}$ is related to its observed brightness temperature, $T_{b,obs}$ through the following well-known expression:
\begin{equation}\label{brightnesstemp}
\theta_{src}=\sqrt{\dfrac{c^{2}S_{\nu}}{2 \pi \nu^{2} kT_{b,obs}}},
\end{equation}
where $\nu$ is the observing frequency, $S_{\nu}$ is the observed flux density, $c$ is the speed of light and $k$ is the Boltzmann constant. On cosmological scales, the observed brightness temperature is a factor of $(1+z)$ lower than the brightness temperature in the frame of emission, $T_{b,em}$, due to the expansion of the Universe. $T_{b,em}$ is in turn a function of the intrinsic brightness temperature of the source, $T_{b,int}$, Doppler-boosted by a factor $\delta$ due to the effects of relativistic beaming in AGN jets. Since only the most compact components of the source core scintillate in the presence of turbulence in the ISM, and it is the sizes of these compact regions that we are inferring from the source variability, we multiply $S_{\nu}$ by the compact fraction, $f_{c}$ of the source. Equation~\ref{brightnesstemp} then becomes:
\begin{equation}\label{brightnesstemp2}
\theta_{src}=\sqrt{\dfrac{(1+z) c^{2} f_{c} S_{\nu}}{2 \pi \nu^{2} k \delta T_{b,int}}},
\end{equation}
This is the angular diameter of the source as it appears to the scattering screen in the ISM, assuming no additional increase in apparent size due to extrinsic propagation effects. Therefore, $\theta_{src} \propto (1+z)^{0.5}$ and we can expect the ISS amplitudes to decrease with redshift, contingent upon the following assumptions:
\begin{enumerate}
\item The mean flux densities of the sources do not vary with redshift, i.e. the sample is flux-limited. 

\item The intrinsic brightness temperatures, $T_{b,int}$, have a cutoff, either at the inverse Compton limit at $\sim 10^{12}$ K \citep{kellermannpauliny-toth69}, or at the energy equipartition limit at $\sim 10^{11}$ K \citep{readhead94,lahteenmakietal99}.

\item The mean Doppler boosting factors and compact fractions of the sources remain constant and do not evolve with redshift.

\item Any possible effect of decreasing angular sizes of the optically thick cores with increasing rest-frame emission frequencies is ignored. 
\end{enumerate}

Based on this model, we performed numerical computations using the fitting function in \citet{goodmannarayan06}, plotting $D(\rm 4d)$ of the weak and strong sources against redshift for various values of the Doppler boosting factor in Figure~\ref{redshiftobmodel}. We used the fiducial values shown in Table~\ref{inputparam1} as the model parameters. We assumed that the phase fluctuations at the scattering screen obey a power law with a Kolmogorov spectrum, and that the sources have a Gaussian intensity profile. The mean $D({\rm 4d})$ obtained from the observations are shown for two redshift bins, separating the weak ($S_{4.9} < 0.3$ Jy) and strong ($S_{4.9} \geq 0.3$ Jy) sources. 

The observed redshift dependence of the mean values of $D({\rm 4d})$ in the 102 $-0.4 < \alpha_{4.9}^{8.4} < 0.4$ sources appears to be consistent with the model and its assumptions. Even with the possible bias towards more variable sources at $z < 2$, the decrease in mean ISS amplitudes is no greater than that expected from this model. As further confirmation, this agreement holds, within the $1\sigma$ error bars, even when only the 72 $>3\sigma$ variable sources were used in obtaining the mean values of $D({\rm 4d})$. Looking at Figure~\ref{redshiftobmodel}, we see that the model successfully explains the weaker redshift dependence of $D_{8.4}({\rm 4d})$ as compared to that of $D_{4.9}({\rm 4d})$, without having to invoke scatter broadening in the IGM. Furthermore, it explains why the ISS amplitudes of the strong sources have a weaker redshift dependence than that of the weak sources, as also reported in Pursimo et al. (submitted) for the larger MASIV sample.

The observations and assumptions of the model, particularly that of constant mean Doppler boosting factors with increasing redshift, are also consistent with other statistical studies of Doppler-boosted AGN sources. Monte Carlo simulations by \citet{listermarscher97} and recent observational data \citep{hovattaetal09} suggest that the mean Doppler factors for a flux-limited sample of sources remains relatively constant at $z > 0.6$. The highly beamed sources with large Doppler factors are very rare, since their jets need to be aligned very close to the line of sight to be detectable. Considering that these large $\delta$ sources would also appear very luminous, one would expect to detect more of these sources with increasing redshift where the volume is also larger (thus increasing the likelihood of detecting these rare sources). However, this is offset by the large range in intrinsic luminosities of the sources, so that sources with large intrinsic luminosities and low Doppler factors (due to large angles of orientation) will also be included in the high-redshift sample. While the range of source Doppler factors increases with redshift in a flux-limited sample, the mean remains the same. According to \citet{listermarscher97} and \citet{arshakianetal10}, at redshifts $z < 0.6$, the mean Doppler boosting factor is lower due to the scarcity of sources with large Doppler factors ($\delta > 20$) within the small volume at such redshifts. This may explain why the most variable sources are not found below $z < 0.6$ at both frequencies, as can be seen in the scatter plots of Figure~\ref{redshiftobmodel}. This effect is seen in the original MASIV data as well (Figure 13 in \citet{lovelletal08}).

The marginal increase in $R_{D}$ with redshift as seen in Figure~\ref{sfxsfcredshift} and Table~\ref{variousRDz} can also be sufficiently explained with the same model of decreasing $T_{b,obs}$ with redshift. Figure~\ref{sfxsfcredshiftbinmod} shows model values of $R_{D}$ (blue and red curves), calculated using the \citet{goodmannarayan06} fitting formula and the same input parameters in Table~\ref{inputparam1}. The binned plots depict the fitted $R_{D}$ at low and high redshifts, with the error bars given by the 68\% confidence bounds (roughly equivalent to $1\sigma$ errors). The model calculations show that $R_{D}$ can indeed increase with redshift without including any scatter broadening effects. This redshift dependence of $R_{D}$ arises due to the increase in source angular size with redshift, which in turn increases the timescale of the scintillations for a fixed scattering screen velocity. As explained in Section~\ref{ratios}, the timescales can increase sufficiently so that the SFs do not saturate within four days, leading to higher values of $R_{D}$. This also explains why the sources with high $R_{D}$ also tend to be strong sources with larger angular sizes rather than weak sources, as seen for the fitted $R_{D}$ from the observations. To further illustrate this point, we include in Figure~\ref{sfxsfcredshiftbinmod} the model values of $R_{D}$ for the case where the scattering screen velocity is reduced to 20 ${\rm kms^{-1}}$ for the strong sources, shown as black curves. In this case, the SFs take a longer time to saturate, thereby increasing $R_{D}$ even further. While the increase in fitted $R_{D}$ from low to high redshift appears larger than that predicted by the model for the weak sources, this difference is $\lesssim 2\sigma$. In any case, this difference cannot be attributed to scatter broadening since the fitted $R_{D}$ of the strong sources is clearly larger than that of the weak sources at both low and high redshift.   

We applied the Monte Carlo method to the \citet{goodmannarayan06} fitting functions to obtain simulated distributions of $D_{4.9}({\rm 4d})$, $D_{8.4}({\rm 4d})$ and $R_{D}$ of each source for further comparisons. Figure~\ref{montecarlocorr} shows observed values of $D_{4.9}({\rm 4d})$, $D_{8.4}({\rm 4d})$ and $R_{D}$ of the $-0.4 < \alpha_{4.9}^{8.4} < 0.4$ sources (72 sources with $\gtrsim 3\sigma$ variability in the case of $R_{D}$) plotted against their corresponding median simulated values. The horizontal error bars represent the median absolute deviations of the simulated values of $D_{4.9}({\rm 4d})$, $D_{8.4}({\rm 4d})$ and $R_{D}$ for each source.
   
For each source, the 4.9 GHz and 8.4 GHz mean flux densities, as well as source redshift, are kept constant at their observed values, since these parameters are known. We also keep the transition frequency and scattering screen distance of each source constant, estimated from the line-of-sight H$\alpha$ intensity and Galactic latitude of the source (see Appendix~\ref{appendixa} for details). We fixed the intrinsic brightness temperatures and scattering screen velocities at the typical values of $10^{11}$ K and 50 ${\rm kms}^{-1}$ respectively for all sources. We then randomly generated 1000 values of the Doppler boosting factor and source compact fraction for each of the sources, with Gaussian distributions peaked at 15 and 0.5, and standard deviations of 4 and 0.1 respectively. 

The observed values of ${\rm log_{10}}[D_{4.9}({\rm 4d})]$, ${\rm log_{10}}[D_{8.4}({\rm 4d})]$ and ${\rm log_{10}}[R_{D}]$ show statistically significant correlations with their simulated counterparts. We obtained Pearsons linear correlation coefficients of 0.54 for both ${\rm log_{10}}[D_{4.9}({\rm 4d})]$ and ${\rm log_{10}}[D_{8.4}({\rm 4d})]$ respectively, with probabilities of $4.0 \times 10^{-9}$ and $3.9 \times 10^{-9}$ that they were obtained by chance. We obtained a weaker correlation coefficient of 0.23 for log$[R_{D}]$, with a probability of 0.05 that this was obtained by chance. The correlation is weaker for $R_{D}$ due to the presence of sources with observed $R_{D}$ values below 0.4, possibly due to errors in the estimation of $R_{D}$ or intrinsic source sizes with frequency dependences flatter than the $\nu^{-1}$ used in our model. 

Our simulated dataset exhibits similar trends to that of the observations. K-S tests performed on the simulated SFs show that $D_{4.9}({\rm 4d})$ and $D_{8.4}({\rm 4d})$ of the $z < 2$ sources are larger than that of the $z > 2$ sources, with probabilities of $6.1\times 10^{-5}$ and $9.0 \times 10^{-5}$ that these were obtained by chance. As in our observational data, the high-redshift $R_{D}$ values are not significantly larger than the low-redshift $R_{D}$ values, even though the six sources with the largest simulated $R_{D}$ values are all at $z > 2$ (Figure~\ref{montecarlocorr}). Additionally, the K-S tests show that $R_{D}$ values for both the $S_{4.9} \gtrsim 0.3$ Jy and $I_{\alpha} \gtrsim 5.0$ R source samples are larger than $R_{D}$ values in the $S_{4.9} < 0.3$ Jy (with a probability of $2.8 \times 10^{-4}$ that this result was obtained by chance) and $I_{\alpha} < 5.0$ R (with a probability of $1.1 \times 10^{-3}$ that this result was obtained by chance) source samples, broadly consistent with that of our observations.

This model of $D(\tau) \propto (1+z)^{0.5}$ is consistent with that of the original 4.9 GHz MASIV dataset as well, for $\sim 250$ sources in which redshift data are available, except at $z > 3$ where the observed $D({\rm 2d})$ is $\sim 2\sigma$ below the predicted curves (Pursimo et al., submitted). However, this steeper than predicted drop in $D({\rm 2d})$ can be attributed to the additional effect of the $z$-$\alpha$ correlation, which cannot be further analyzed for the original MASIV sample due to observations at only a single frequency. The model and Monte Carlo simulations will hence need to be applied to the full MASIV dataset to further test this model of decreasing observed brightness temperatures, pending the acquisition of optical spectroscopic redshifts for the remaining sources for which they are not available. The larger sample may also allow us to break the redshift-luminosity degeneracy, providing a stronger test of whether we are observing a redshift cutoff or luminosity cutoff in $D(\tau)$.

       \subsection{The $z$-$\alpha_{4.9}^{8.4}$ Correlation: Selection Effect or Source Evolution?}\label{z-specindex}
         We discuss why the mean source spectral indices steepen with increasing redshift (Figure~\ref{spec_vs_z_before}), and why 13 out of the 15 sources with $\alpha_{4.9}^{8.4} < -0.4$ in the sample lie at $z > 2$. This effect partially accounts for the redshift dependence of ISS in our sample, as it has been established in Paper I that the mean $D({\rm 4d})$ of the $\alpha_{4.9}^{8.4} < -0.4$ sources is a factor of $\sim 6$ lower than that of the $\alpha_{4.9}^{8.4} \geq -0.4$ sources. The $\alpha_{4.9}^{8.4} < -0.4$ sources therefore either have source sizes that are on average a factor of $\sim 2$ larger or compact fractions that are a factor of $\sim 2.5$ lower than their $\alpha_{4.9}^{8.4} \geq -0.4$ counterparts.  
 
This $z$-$\alpha^{4.9}_{8.4}$ correlation could simply be a selection effect. As mentioned in Section~\ref{selection}, the original MASIV Survey sources were selected to have $\alpha^{1.4}_{8.4} > -0.3$, but the sample could have been contaminated by sources whose spectral indices were inaccurately estimated due to the non-coeval, variable flux densities. The comoving spatial density of flat-spectrum radio sources has been found to decrease by a factor of $\sim 5$ between redshifts $2 < z < 4$ \citep{peacock85, dunloppeacock90}, perceived as evidence for a peak in quasar activity at $z \sim 2.5$. A similar but slower decline was found for steep-spectrum sources \citep{dunloppeacock90}, which may explain why there are more of these steeper $\alpha_{4.9}^{8.4}$ sources at high redshift. Two of the high-redshift, steep-spectrum sources have been optically identified as narrow-line radio galaxies (Type 2 AGNs with narrow emission lines), and so may be classical steep-spectrum sources. After removing these two sources from the original sample of 128 sources, the K-S Test still finds a statistically significant difference in the spectral indices in the low and high-redshift source samples. The fact that the $z < 2$ sample is biased towards variable sources, while all sources with known redshifts at $z > 2$ were selected regardless of variability, could also have biased the low-redshift sample towards flatter $\alpha_{4.9}^{8.4}$, and vice versa for the high-redshift sample. 

It is also possible that some of these sources have convex spectra that steepen at higher frequencies, as found in gigahertz peaked-spectrum (GPS) sources; k-correction effects then lead to a steepening of spectral indices due to increasing rest-frame emission frequencies with increasing redshift. \citet{jarvisrawlings00} have demonstrated that a significant portion of the most luminous radio-selected flat-spectrum sources are GPS sources. Since these GPS sources are not Doppler-boosted, they are less compact and therefore scintillate less. Furthermore, \citet{chhetrietal12} recently found a steepening of source spectral indices with redshift in compact radio sources from the Australia Telescope 20 GHz (AT20G) Survey. This $z$-$\alpha$ correlation was discovered to be more prominent when $\alpha_{4.8}^{8.6}$ is used as compared to $\alpha_{1.0}^{4.8}$, which they attribute to spectral curvature and k-correction effects. 

A natural and physical explanation for this $z$-$\alpha_{4.8}^{8.6}$ correlation is also conceivable. The steepening of source spectral indices with redshift has long been observed in radio galaxies (classical steep-spectrum sources) \citep{laingpeacock80, macklin82}, where spectral index cut-offs have been successfully used to search for radio galaxies at the highest redshifts. Traditional explanations for this correlation include (1) k-correction effects in sources with convex spectral energy densities \citep{gopal-krishna88} (2) inverse Compton losses associated with Cosmic Microwave Background (CMB) photons whose energy densities scale with $(1+z)^{4}$ \citep{krolikchen91} and (3) a luminosity-spectral index relation coupled with the Malmquist bias \citep{laingpeacock80, blundelletal99}. More recently, \citet{klameretal06} argue that this $z$-$\alpha$ correlation could be due to higher ambient densities at high redshift, resulting in increased synchrotron and inverse Compton losses in pressure-confined radio lobes. The evidence comes from the observed properties of high-redshift radio galaxies, which include (1) similarities to low-redshift radio galaxies residing in dense clusters, (2) extreme rotation measures (RMs), and (3) knotty radio emission interpreted as frustrated jets in dense and clumpy environments. For radio galaxies, one would expect to observe increasingly compact sources at high redshift. All known radio-loud AGNs at $z > 4$ are steep spectrum sources, with VLBI images revealing compact double structures reminiscent of compact steep-spectrum (CSS) and GPS sources \citep{freyetal08, freyetal10, freyetal11}, thought to be young and `frustrated' radio galaxies. 

If the flat-spectrum, Doppler-boosted population of radio-loud AGNs reside in similar environments at high redshift, it is likely that pressure from the surrounding IGM, whose densities scale with $(1+z)^{3}$, will reduce their Doppler boosting factors. This in turn reduces the compact fractions of the sources and reduces their scintillation amplitudes. This will also result in a steepening of spectral indices, as the contribution of the optically thick core components to the mean observed flux densities is reduced relative to that of the optically thin jet components. 

Multifrequency VLBI studies based on new observations and archival data will be needed to determine if this $z$-$\alpha_{4.9}^{8.4}$ correlation and its relationship to the redshift dependence of ISS, is mainly due to selection effects or interesting physical phenomena related to AGN evolution.

      \subsection{Constraints on IGM Scattering and Turbulence}\label{constraints}
         While our observations provide no clear detection of scatter broadening in the IGM or that it has any significant redshift dependence between the $z < 2$ and $z > 2$ subsamples, we can place strong constraints on it. The top panel of Figure~\ref{IGMscatsky} shows estimated apparent angular sizes of all sources at 4.9 GHz, $\theta_{4.9}$, calculated from their $D_{4.9}({\rm 4d})$ values using the \citet{goodmannarayan06} fitting functions. For all sources in which $D_{4.9}({\rm 4d}) \geq D_{noise}$, the upper limits of $\theta_{4.9}$ are calculated using $f_{c} \sim 1$, while the lower limits are calculated using $f_{c} \sim 0.1$. For sources in which $D_{4.9}({\rm 4d}) < D_{noise}$, upper limits to $\theta_{4.9}$ cannot be obtained, while the lower limits are calculated using $f_{c} \sim 0.1$ and setting $D_{noise}$ as the upper limit to the source variability. For each source, we used $\nu_t$ and $D_{\rm ISM}$ calculated in Appendix~\ref{appendixa}, which were also used in the Monte Carlo simulations described in Section~\ref{expansion}. The estimation of $\theta_{4.9}$ makes no assumptions about the brightness temperatures and Doppler boosting factors of the sources. The upper limits of $\theta_{4.9}$ are also effectively upper limits of the 4.9 GHz $\theta_{scat}$ and $\theta_{src}$ for the sight-lines to our sources (Equation~\ref{theta}); they are shown proportional to the sizes of the circles in the lower panel of Figure~\ref{IGMscatsky} in Galactic coordinates. 

We go one step further in constraining IGM scatter broadening by making use of the fitted $R_{D}$ obtained for the $\gtrsim 3\sigma$ variable sources. At 4.9 GHz, the upper limit to scatter broadening can be formulated from Equations~\ref{RDweak} and \ref{theta} in the weak ISS regime as: 
\begin{equation}\label{igmscat}
\theta_{scat} \leq \theta_{src(max)}{\left[ \frac{2.16 {\left( R_{D(max)}\right)}^{\frac{6}{7}}  - 1}{1-0.59{\left( R_{D(max)}\right)}^{\frac{6}{7}} }\right]}^{0.5},
\end{equation}
for $0.4 < R_{D(max)} < 1.8$, where $R_{D(max)}$ is the upper limit of $R_{D}$ and $\theta_{src(max)}$ is the upper limit to the intrinsic source angular size at 4.9 GHz. Again, we made use of the relations $\theta_{src} \propto \nu^{-1}$ and $\theta_{scat} \propto \nu^{-2.2}$. This inequality posits that sources scintillating in the weak ISS regime will have $R_{D} \sim 0.4$ when completely dominated by intrinsic source size effects. As an increase in IGM scatter broadening increases $R_{D}$, the upper limit to $\theta_{scat}$ is determined by the level of increase in $R_{D}$ above this nominal value. The dominance of $\theta_{scat}$ is also dependent on $\theta_{src}$; sources with smaller intrinsic angular sizes are more likely to be dominated by $\theta_{scat}$ than sources with larger angular sizes.

We use the weak ISS approximation by \citet{narayan92} and \citet{walker98} here because it gives the most conservative upper limit to scatter broadening in the IGM and is not dependent on any other model parameters. Comparing the \citet{goodmannarayan06} fitting function with the weak ISS model \citep{narayan92,walker98} in calculating $R_{D}$, we found that both provide similar constraints when $\theta = \sqrt{\theta_{src}^{2} + \theta_{scat}^{2}}$ is $\lesssim 50 \mu$as (see Figure~\ref{igmconstraints}). As opposed to the weak ISS model where $D({\rm 4d}) \approx 2m^{2}$ is assumed, $R_{D}$ rises faster in the \citet{goodmannarayan06} fitting function with increasing $\theta$ when $\theta$ is of an order $\sim 100 \mu$as, as the scintillation timescales exceed 4 days and the SFs do not saturate (the same reason why $R_{D}$ increases marginally with redshift in Figure~\ref{sfxsfcredshiftbinmod}). Although the \citet{goodmannarayan06} fitting functions give stronger constraints, they are very dependent on the parameters of the model when $\theta$ is large. 
 
Since the upper limit of $R_{D}$ for the low-redshift sources is 0.57 at 99\% confidence, we obtain $\theta_{scat} \lesssim 110 \mu$as at 4.9 GHz, assuming that the intrinsic angular sizes of the scintillating components in all our sources are $\lesssim 150 \mu$as (as seen for the $\gtrsim 3\sigma$ variable sources at low redshift in the upper panel of Figure~\ref{IGMscatsky}). In discussing the feasibility of using the Square Kilometre Array to detect intergalactic scatter broadening, \citet{lazioetal04} and \citet{godfreyetal11} propose that angular resolutions better than 4 mas at 1.4 GHz and 80 mas at 0.33 GHz are required. Our most conservative constraints push these limits lower. A simple extrapolation gives $\theta_{scat} \lesssim 1.7$ mas at 1.4 GHz and $\theta_{scat} \lesssim 42$ mas at 0.33 GHz.  

The strongest constraints can be derived from the $R_{D}$ of the weak sources, which is also no more than 0.57 at 99\% confidence for the weak, low redshift sources. With flux densities of $\sim 0.1$ Jy, $\theta_{src}$ can be as low as $\sim 10 \mu$as. For example, the source PKS 1519-273 (not in our sample) has an estimated core size as low as 15 to 30 $\mu$as \citep{macquartetal00}, while the most compact component of the extreme scintillator J1819+3845 has been estimated to be as small as $\sim 7 \mu$as \citep{macquartdebruyn07}. In our present data, the rapid scintillator J1328+6221 \citep{koayetal11b} has the lowest upper limit of $\theta_{4.9}$, estimated to be $\lesssim 15 \mu$as. Although Figure~\ref{IGMscatsky} shows that the estimated lower limits of $\theta_{4.9}$ in some sources drop well below $1\mu$as if their compact fractions are suffciently small, it is unknown if the compact fractions of these sources do indeed have values as low as 0.1. 

The very compact, $\sim 10 \mu$as sources give $\theta_{scat} \lesssim 8 \mu$as at 4.9 GHz. Again, this can be extrapolated to $\theta_{scat} \lesssim 126 \mu$as at 1.4 GHz and $\theta_{scat} \lesssim 3$ mas at 0.33 GHz. Also, $\theta_{scat} \lesssim 264 \mu$as at 1.0 GHz, which is about a factor of 2 lower than the upper limit of $500 \mu$as obtained by \citet{lazioetal08} at the same frequency. This limit is comparable to that for the sight-line to the Gamma-ray Burst GRB 970508 inferred from its angular size of $\lesssim 3 \mu$as at 8.4 GHz ($\lesssim 9 \mu$as at 4.9 GHz), determined from observations of diffractive scintillation in its radio afterglow \citep{frailetal97}. 
 
A common parameter used to quantify the level of turbulence in the ISM is the spectral coefficient, $C_n^2$, for a truncated power law distribution of electron density fluctuations ($\delta n_{e}$) in the ISM:
\begin{equation}\label{powerlaw}
P_{\delta n_{e}}(q) = {C_n^2}q^{-\beta}, \; \frac{2\pi}{l_0} \lesssim q \lesssim \frac{2\pi}{l_1}.
\end{equation}
$q$ is the wavenumber, $l_0$ and $l_1$ are the outer and inner scales of $\delta n_{e}$, and $\beta$ is usually assumed to be $11/3$ for a Kolmogorov spectrum. The scattering measure (SM), which can be derived from observables, is then the line-of-sight path integral of $C_n^2$ to the source at distance $D_S$:
\begin{equation}\label{SMdefinition}
{\rm SM} = \int_0^{D_S} ds C_n^2.
\end{equation}

Maximum values of the SM for the IGM can be computed from the upper limits of $\theta_{scat}$, using the equation in \citet{taylorcordes93}, extended to cosmological scales (Koay \& Macquart, in prep):
\begin{equation}\label{SM}
{\rm SM} \lesssim {\left[ \frac{\theta_{scat(max)}}{128 \,{\rm mas}} \left( \frac{D_S}{D_{LS}} \right)
{\left( \frac{\nu}{{\rm 1 \, GHz}}\right)}^{2.2}(1 + z_{L})^{1.2}\right]}^{\frac{5}{3}} {\rm kpc \, m^{-\frac{20}{3}}}
\end{equation}
where in the cosmological context, $D_{S}$ is the angular diameter distance to the source and $D_{LS}$ is the angular diameter distance from the source to the scattering screen in the IGM. A Kolmogorov spectrum is assumed for the electron density fluctuations in the IGM, modeled as a thin screen located at an effective distance equivalent to a redshift of $z_{L}$. It is likely that scatter broadening will be dominated by nearby screens, due to the geometrical `lever arm' effect (see \citet{rickettetal07} for more details), as well as the redshift dependence of the rest-frame frequency at the scattering screen for a fixed observing frequency. This may explain why we did not observe a significant redshift dependence of $R_{D}$, even though the mean baryonic density of the Universe scales with $(1+z)^3$, and the probability of sight-lines intersecting potential scattering regions (i.e. galaxies and their progenitors, the intracluster medium, void walls) increases with source redshift. At $z_{L} \sim 0$, we obtain SM $\lesssim 3.3 \times 10^{-5}\, {\rm kpc \, m^{-20/3}}$ for $\theta_{scat(max)} \sim 8 \mu$as. 

We can also place quantitative limits on the level of turbulence in the IGM. The SM can be expressed as \citep{lazioetal08}:
\begin{equation}\label{SM2}
{\rm SM} = C_{\rm SM}\overline{F{n_e^2}}D_{S},
\end{equation}
where the constant $C_{\rm SM} = 1.8 {\rm m}^{-20/3} \, {\rm cm}^6$, $n_e$ is the electron density and $F$ is a fluctuation parameter, given by \citep{taylorcordes93}:
\begin{equation}\label{F}
F = \frac{\zeta \epsilon^{2}}{\eta} {\left( \frac{l_{0}}{1 {\rm pc}}\right)}^{-\frac{2}{3}}. 
\end{equation}
$\zeta$ is the normalized intercloud variance of the mean electron densities of each cloud, $\epsilon$ is the normalized variance of the electron densities within a single scattering cloud, $\eta$ is the filling factor for ionized clouds in the path, and $l_{0}$ is the outer scale of the density fluctuations with Kolmogorov turbulence. Following \citet{lazioetal08}, we adopt $\overline{n_e} < 2.2 \times 10^{-7} {\rm cm}^{-3}$ at $z \sim 0$ based on the assumption that helium is fully ionized, and $\zeta \sim \epsilon \sim \eta \sim 1$. We therefore obtain $F \lesssim 230$ and $l_0 \gtrsim 3 \times 10^{-4}$ pc for a source at $z \sim 1$ ($D_S \sim 1.7$ Gpc). This is consistent with what we know, where $l_0$ can range from $\sim 1$ pc if the scattering occurs at an intervening spiral galaxy similar to our own, up to the $\sim$ 0.1 Mpc scales of the largest jet sources that can inject turbulence into the IGM. We also note that \citet{cordeslazio03} give $F \sim 0.2$ and $F \sim 10$ respectively in the thick disk and spiral arm components of our Galaxy for comparisons.

   \section{Conclusions}\label{conclusion}
         We analyzed data from a VLA survey of ISS in 128 sources at 4.9 GHz and 8.4 GHz to determine the origin of the redshift dependence of AGN ISS. We made use of two ISS models to interpret the data, one an analytical approximation \citep{narayan92,walker98} for the asymptotically weak and strong ISS regimes, the other a fitting function \citep{goodmannarayan06} that is applicable at the transition between the weak and strong ISS regimes. We also took into consideration the various selection effects in the source sample. We can summarize our findings as follows:
\begin{enumerate}
\item The examination of the ratio of the SFs for each individual source is a good strategy for mitigating source selection effects in the sample, negating the redshift dependence of source luminosities, compact fractions, Doppler boosting factors and rest frame emission frequencies. Three effects can increase the ratio of $D_{8.4}({\rm 4d})$ to $D_{4.9}({\rm 4d})$ from $R_{D} \sim 0.4$ in the weak ISS regime to $R_{D} \gtrsim 1$: (1) scatter broadening, (2) scintillation at the strong ISS regime, or at the transition between weak and strong ISS, (3) and sufficiently large scintillation timescales so that the SFs do not saturate at one or both frequencies so that $D_{8.4}({\rm 4d})$ rises faster than $D_{4.9}({\rm 4d})$.

\item The examination of the correlation of the SF ratios, $R_{D}$, with source mean flux densities, line-of-sight H$\alpha$ intensities and source redshifts allow these three competing causes of large $R_{D}$ to be discriminated. 

\item We observed no significant scatter broadening in our sources at the scales of tens and hundreds of $\mu$as probed by our survey, due either to the ISM or the IGM. We found no significant increase of IGM scatter broadening in the $z > 2$ sources compared to that of the $z < 2$ sources, ruling it out as the cause of the redshift dependence of ISS. In performing the analysis of $R_{D}$ for the $\geq 3 \sigma$ variable sources, we note that we are including only the most variable sources at any redshift, which could mean that they are the least scatter broadened sources. Similar observations with higher sensitivity instruments such as the planned Square Kilometre Array (SKA), will enable $R_{D}$ to be accurately estimated for sources with even lower $D({\tau})$ at both frequencies, to determine if the sources with lower variability amplitudes are scatter broadened. Another weakness of the present analyses is that $\sim 85\%$ of the $-0.4 < \alpha_{4.9}^{8.4} < 0.4$, high-redshift sources lie between $2 < z < 3$. There is a dearth of sources at $z > 3$. The inclusion of more $z > 3$ sources in similar future surveys will more robustly determine if there is significant scatter broadening of sources beyond $z \sim 3$.

\item We infer that angular broadening in the IGM at 4.9 GHz is $\lesssim 110 \mu$as for all lines of sight to our sources, and down to $\lesssim 8 \mu$as for sight-lines to the $\sim 10 \mu$as sources. We also obtain an upper limit to the scattering measure (SM) of the IGM at $3.3 \times 10^{-5}\, {\rm kpc \, m^{-20/3}}$ for these latter lines of sight. 

\item We found a statistically significant steepening of source spectral indices ($\alpha_{4.9}^{8.4}$) with source redshift, which partially accounts for the redshift dependence of AGN ISS. This $z$-$\alpha_{4.9}^{8.4}$ correlation can be attributed to selection effects or frustrated AGN jets in high-redshift environments. Follow-up high-resolution imaging of these sources using VLBI or space VLBI may help to discriminate between these two effects.

\item Selecting sources in the spectral index range of $-0.4 < \alpha_{4.9}^{8.4} < 0.4$, the redshift dependence of AGN ISS is found to be still significant, and can be successfully modeled by a $(1 + z)^{0.5}$ scaling of intrinsic angular sizes of a flux and brightness temperature-limited sample of sources due to the space-time metric of an expanding Universe. 

\end{enumerate}

   \acknowledgments
   		 JYK is supported by the Curtin Strategic International Research Scholarship (CSIRS) provided by Curtin University. BJR thanks the US National Science Foundation (NSF) for partial support under grant AST 05-07713 and for the hospitality of the Cavendish Astrophysics group at Cambridge University. R. Ojha is supported by an appointment to the NASA Postdoctoral Program at the Goddard Space Flight Center, administered by Oak Ridge Associated Universities through a contract with NASA. We all thank the operators and scientific staff at the VLA; in particular we thank Vivek Dhawan for his extensive advice and help during our long sequence of observations. The VLA is part of the National Radio Astronomy Observatory (NRAO), which is a facility of the NSF operated under cooperative agreement by Associated Universities, Inc. This study made use of data from the Wisconsin H-Alpha Mapper (WHAM) northern sky survey, which is funded by the NSF. We used data obtained from the NASA/IPAC Extragalactic Database (NED), which is operated by the Jet Propulsion Laboratory, California Institute of Technology, under contract with NASA, as well as the SIMBAD database, operated at CDS, Strasbourg, France.

   \appendix
   \section{Estimation of Transition Frequencies and Scattering Screen Distances}\label{appendixa}
   This section presents the relations used to estimate the transition frequency ($v_t$) between weak and strong ISS, as well as the scattering screen distance for each source. These  values were used for the Monte Carlo simulations in Section~\ref{expansion} and to obtain estimates of the apparent angular sizes of the sources in Section~\ref{constraints}.

The emission measure (EM) is the integral of the square of the electron density along the path from the observer to the source, and for the ISM is related to the line-of-sight Galactic H$\alpha$ intensity ($I_{\alpha}$) in units of Rayleighs as \citep{haffneretal98}: 
\begin{equation}\label{EMhalpha}
{\rm EM} = 2.75 \,\, T^{0.9}_{4} I_{\alpha}\,\,{\rm cm^{-6}\, pc},
\end{equation}
where $T_{4}$ is the temperature of the ionized cloud in units of $10^{4}$ K, typically $\sim 8000$ K for the warm ionized medium \citep{haffneretal98}. The transition frequency between weak and strong ISS, is then given by \citet{cordeslazio03} as:
\begin{equation}\label{transfreqSM}
v_{t} = 318 \,\,{\rm SM}^{\frac{6}{17}}{(D_{\rm ISM})}^{\frac{5}{17}}\,\,{\rm GHz},
\end{equation}
where $D_{\rm ISM}$ is the effective distance to the ISM scattering screen in units of kpc, while ${\rm SM}$ is the scattering measure of the ISM, which is defined as the path integral of the strength of turbulence in the ISM along the line-of-sight to the source (see Section~\ref{constraints} for more details on the SM), and has units of ${\rm kpc \, m^{-20/3}}$. \citet{cordeslazio03} also give the relation between the SM and the EM, which for a thin screen can be estimated as:
\begin{equation}\label{EMSM}
{\rm EM} = 544.6\, l_0^{2/3} \epsilon^{-2}(1 + \epsilon^2) {\rm SM} \,\,{\rm pc \, cm^{-6}},
\end{equation}
where $\epsilon$ is the normalized variance of the electron densities within the scattering cloud, which we assume to be $\sim 1$. $l_0$ is the outer scale of the turbulence in units of pc, which has been estimated to be $\lesssim 100$ pc \citep{haverkornetal08}. We use 100 pc for our calculations. Combining Equations~\ref{EMhalpha} to \ref{EMSM}, we obtain: 
\begin{equation}\label{vthalpha}
v_{t} = 318 \left({D_{\rm ISM}}\right)^{\frac{5}{17}} {\left[ \frac{I_\alpha}{\rm 198\, R} \left( \frac{T_{4}^{0.9}\epsilon^{2}}{l_0^{2/3}(1 + \epsilon^{2})} \right) \right]}^{\frac{6}{17}} \; {\rm GHz}.
\end{equation}

The distance to the scattering screen, used in Equation~\ref{EMSM} and for the \citet{goodmannarayan06} fitting functions for ISS, is calculated for each source as $D_{\rm ISM} = 0.35 \times {\rm csc}\vert b \vert$ kpc, where $b$ is the Galactic latitude of the source. The value of 0.35 was selected as the constant of proportionality so that $D_{\rm ISM} \sim 0.5$ kpc at $b = \pm 45^{\circ}$, increasing to $\sim 2.0$ kpc at $b = \pm 10^{\circ}$ and decreasing to $\sim 0.35$ kpc at $b = \pm 90^{\circ}$. 

\begin{figure}[!htp]
     \begin{center}
     \includegraphics[scale=0.8]{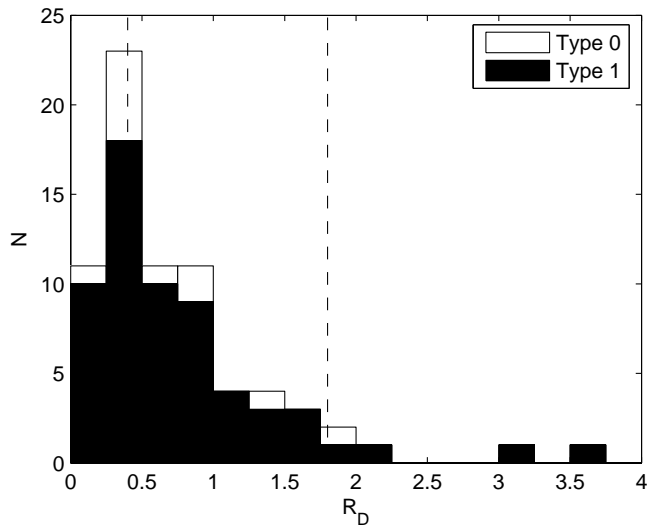}
     \end{center}
     \caption{{Histogram of $R_{D}$ for 72 sources with $D({\rm 4d}) \geq 3\sigma$ at both frequencies, classified as Type 0 or Type 1 AGNs. The dashed vertical lines denote $R_{D} = 0.4$ and $R_{D} = 1.8$.} \label{histoSFxSFc}}
     \end{figure}

\begin{figure}[!htp]
     \begin{center}
     \includegraphics[scale=0.78]{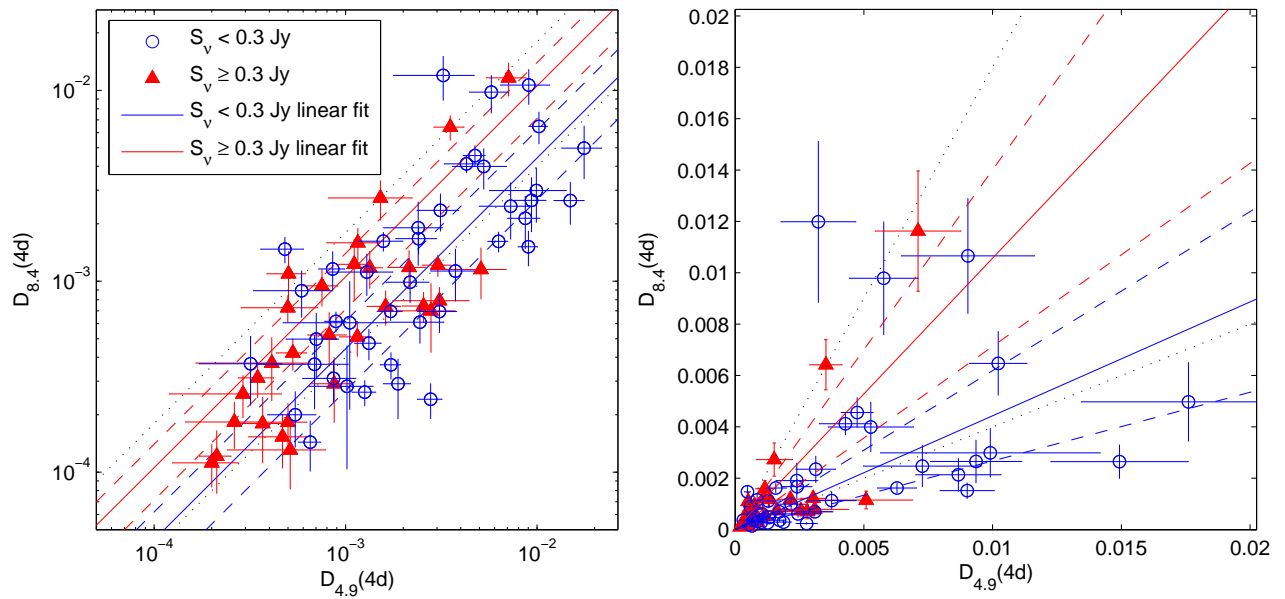}
     \end{center}
     \caption{{Plot of $D_{8.4}({\rm 4d})$ against $D_{4.9}({\rm 4d})$ in logarithmic (left) and linear (right) scales with sources classified as `weak' or `strong' based on their observed mean flux densities at 4.9 GHz, $S_{4.9}$. Only sources with $D({\rm 4d}) \geq 3\sigma$ at both frequencies are included. The dotted lines represent $R_{D} = 0.4$ and $R_{D} = 1.8$. The solid lines represent linear fits to the $S_{4.9} < 0.3$ Jy and $S_{4.9} \geq 0.3$ samples, while the dashed lines represent 99\% confidence bounds for those fits.} \label{sfxsfcweakstrong}}
     \end{figure}

\begin{figure}[!htp]
     \begin{center}
     \includegraphics[scale=0.78]{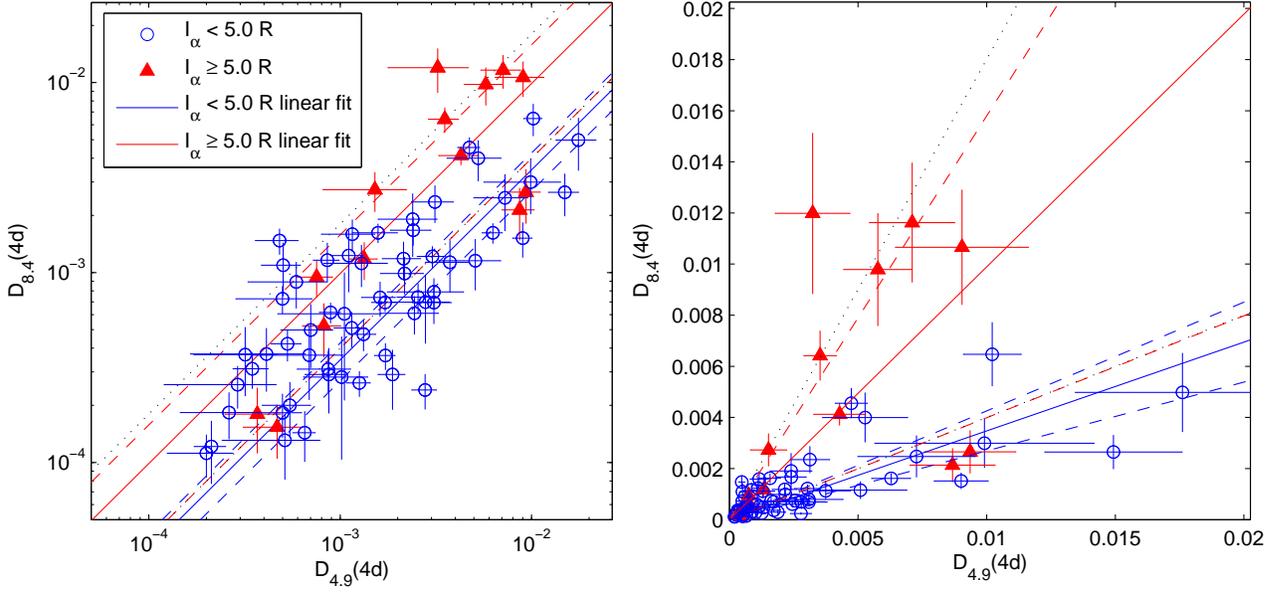}
     \end{center}
     \caption{{Plot of $D_{8.4}({\rm 4d})$ against $D_{4.9}({\rm 4d})$ in logarithmic (left) and linear (right) scales, where the sources are classified based on their line-of-sight H$\alpha$ intensities, $I_{\alpha}$, in units of Rayleighs. Only sources with $D({\rm 4d}) \geq 3\sigma$ at both frequencies are included. The dotted lines represent $R_{D} = 0.4$ and $R_{D} = 1.8$. The solid lines represent linear fits to the $I_{\alpha} < 0.5$ R and $I_{\alpha} \geq 0.5$ R samples, while the dashed lines represent 99\% confidence bounds for those fits.} \label{sfxsfchalpha}}
     \end{figure}

\begin{figure}[!htp]
     \begin{center}
     \includegraphics[scale=0.78]{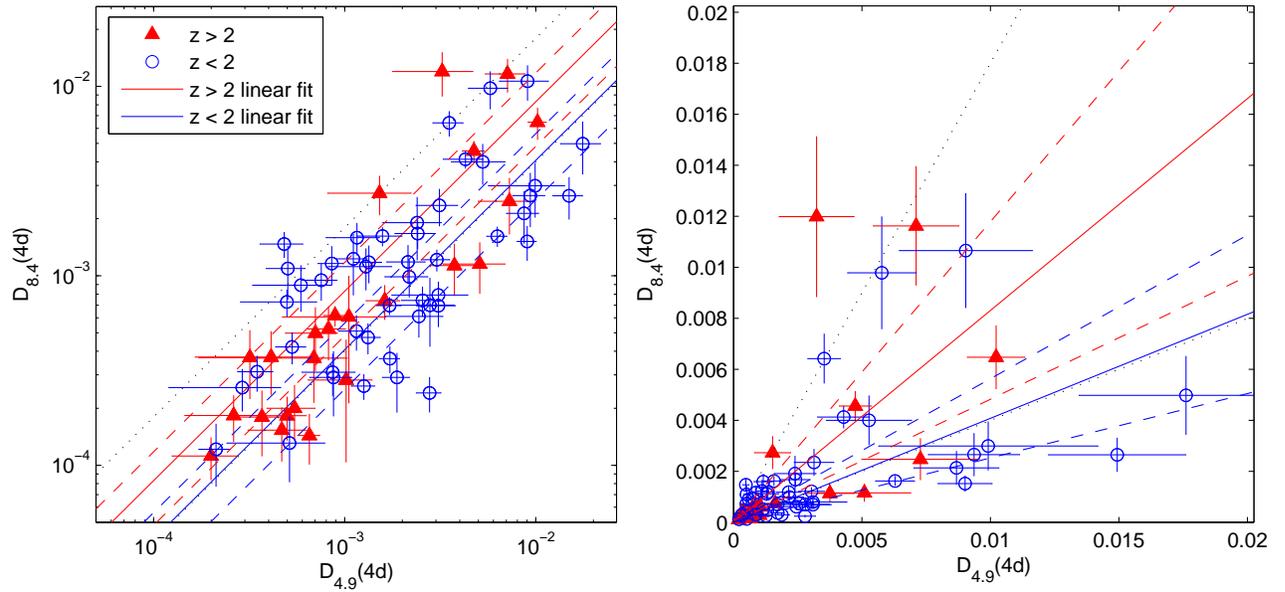}
     \end{center}
     \caption{{Plot of $D_{8.4}({\rm 4d})$ against $D_{4.9}({\rm 4d})$ in logarithmic (left) and linear (right) scales, where the sources are classified based on their redshifts. Only sources with $D({\rm 4d}) \geq 3\sigma$ at both frequencies are included. The dotted lines represent $R_{D} = 0.4$ and $R_{D} = 1.8$. The solid lines represent linear fits to the $z < 2$ and $z > 2$ samples, while the dashed lines represent 99\% confidence bounds for those fits.} \label{sfxsfcredshift}}
     \end{figure}

\begin{figure}[!htp]
     \begin{center}
     \includegraphics[scale=0.82]{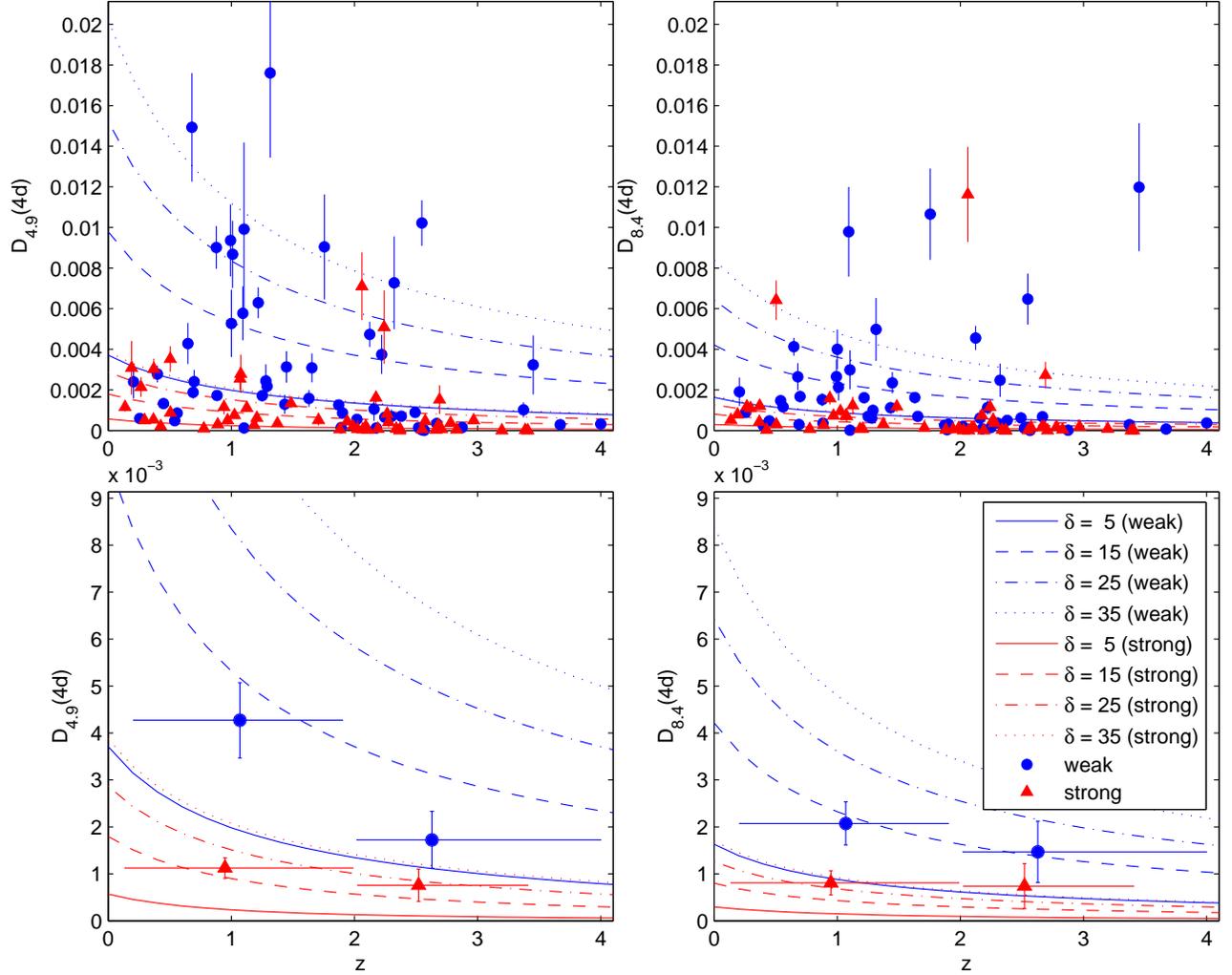}
     \end{center}
     \caption{{Observed $D({\rm 4d})$ at 8.4 GHz (left) and 4.9 GHz (right) plotted against redshift, shown as scatter plots (top) and in bin averages (bottom), for both the weak ($S_{4.9} < 0.3$ Jy) and strong ($S_{4.9} \geq 0.3$ Jy) sources. The vertical error bars in the binned plots represent one standard error in the mean. The lines in all panels show computed model values of $D({\rm 4d})$ for various values of the source Doppler boosting factor, assuming that cosmological expansion leads to a $(1 + z)^{0.5}$ scaling of the intrinsic angular diameter in a flux and brightness temperature-limited sample of sources. The model parameters used are listed in Table~\ref{inputparam1}.} \label{redshiftobmodel}}
     \end{figure}

\begin{figure}[!htp]
     \begin{center}
     \includegraphics[scale=0.8]{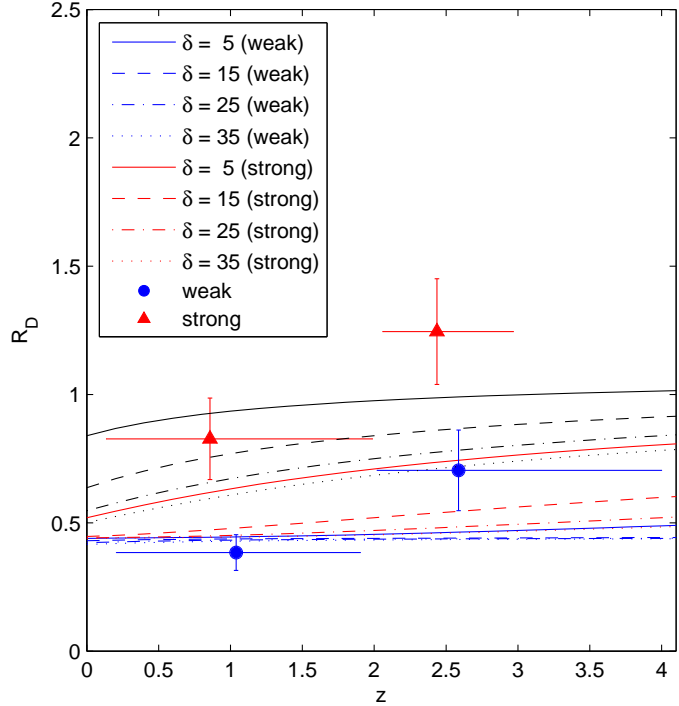}
     \end{center}
     \caption{{Observational values of the fitted $R_{D}$ in two redshift bins with their corresponding 68\% confidence bounds ($\approx 1\sigma$ errors), separated into weak ($S_{4.9} < 0.3$ Jy) and strong ($S_{4.9} \geq 0.3$ Jy) sources. These are shown together with model values of $R_{D}$ for various source Doppler boosting factors. For the blue and red curves, the model parameters are given in Table~\ref{inputparam1}. The black curves show the corresponding model values for the strong sources at a screen velocity of 20 ${\rm kms^{-1}}$ with the other parameters unchanged.} \label{sfxsfcredshiftbinmod}}
     \end{figure}

\begin{figure}[!htp]
     \begin{center}
     \includegraphics[scale=0.65]{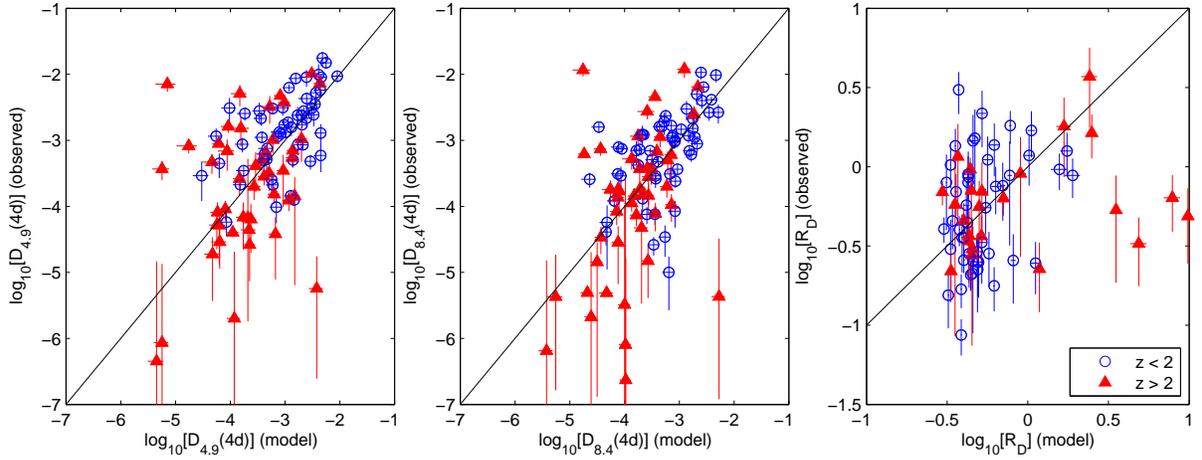}
     \end{center}
     \caption{{Observed values of $D_{4.9}({\rm 4d})$ (left panel), $D_{4.9}({\rm 4d})$ (middle panel) and $R_{D}$ (right panel) plotted against their respective model values obtained by applying the Monte Carlo method to the \citet{goodmannarayan06} fitting functions (described in Section~\ref{expansion}). For $D_{4.9}({\rm 4d})$ and $D_{8.4}({\rm 4d})$, all 102 $-0.4 < \alpha_{4.9}^{8.4} < 0.4$ sources are shown. For $R_{D}$, only the 72 sources with $\gtrsim 3\sigma$ variability are shown. The solid diagonal lines show where the observed values of $D({\rm 4d})$ and $R_{D}$ are equal to their model values.} \label{montecarlocorr}}
     \end{figure}

\begin{figure}[!htp]
     \begin{center}
     \includegraphics[scale=0.7]{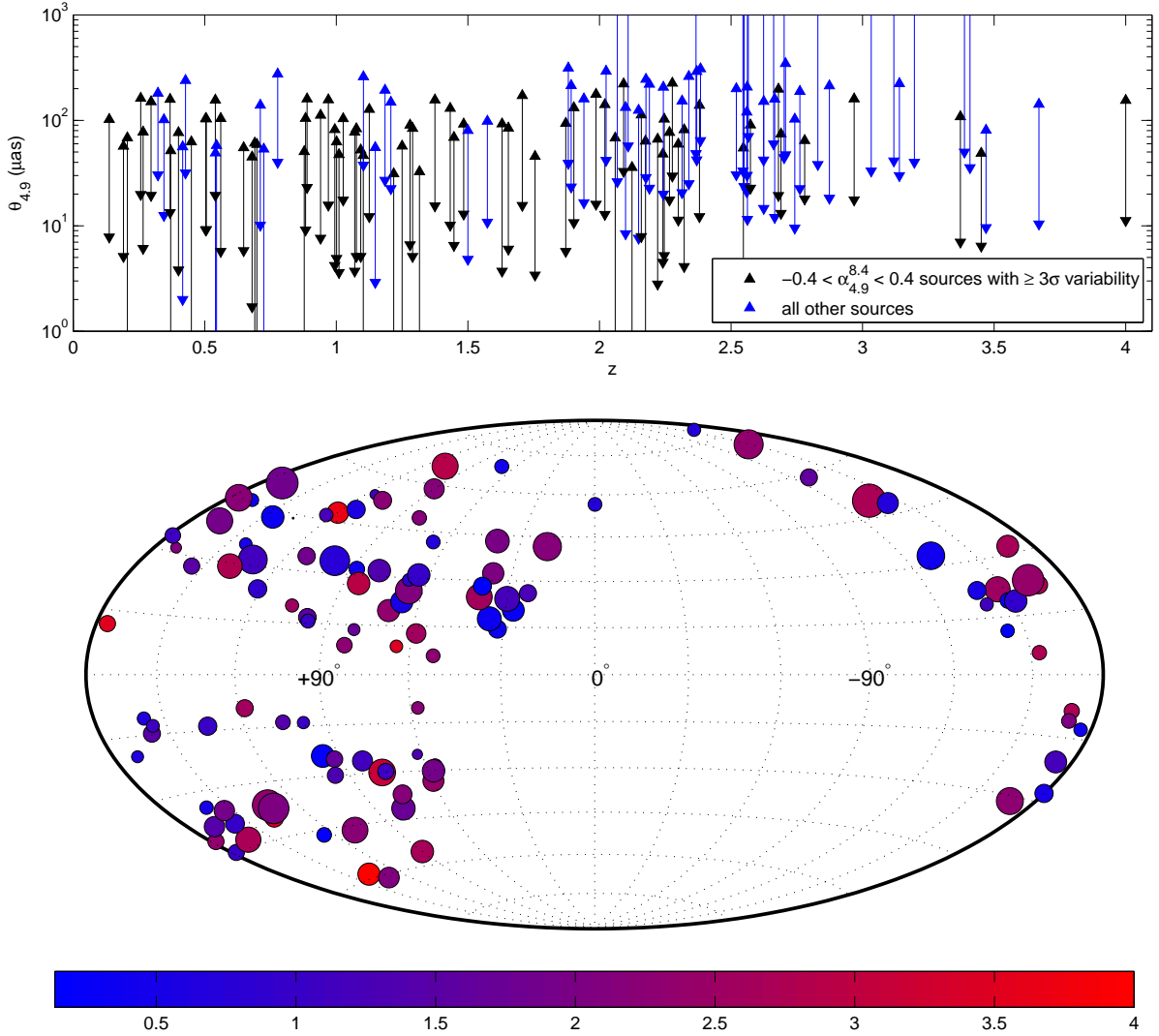}
     \end{center}
     \caption{{Top panel: Constraints on the 4.9 GHz apparent angular sizes, $\theta_{4.9}$, for all 128 sources (including those with $\alpha_{4.9}^{8.4} > 0.4$ and $\alpha_{4.9}^{8.4} < -0.4$), calculated based on $D_{4.9}({\rm 4d})$ using the \citet{goodmannarayan06} fitting function (see Section~\ref{constraints} for more details). Bottom panel: Upper limits of $\theta_{4.9}$ for all sources in which they can be obtained, shown proportional to the sizes of the circles and plotted in Galactic coordinates. They are also effectively upper limits of $\theta_{scat}$ for all lines of sight to our sources. The circles are color-coded based on the redshifts of the sources.} \label{IGMscatsky}}
     \end{figure}

\begin{figure}[!htp]
     \begin{center}
     \includegraphics[scale=0.8]{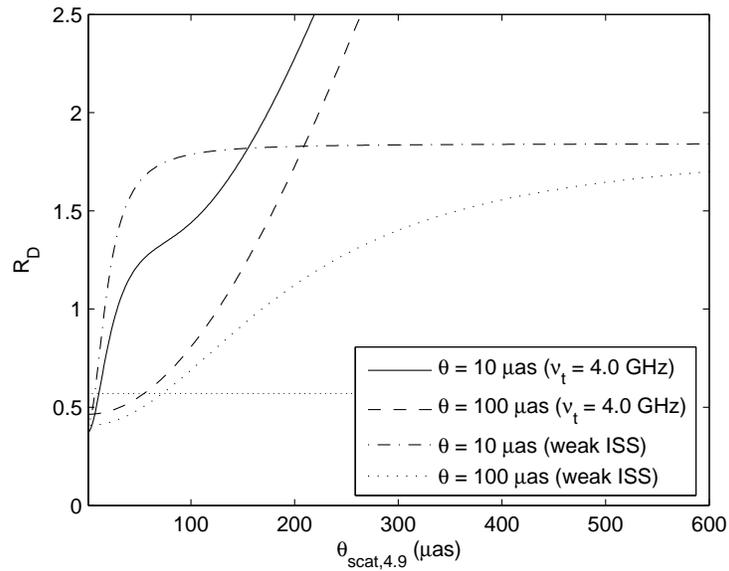}
     \end{center}
     \caption{{$R_{D}$ at increasing values of $\theta_{scat}$ for sources with intrinsic angular sizes of 10 $\mu$as and 100 $\mu$as. The plots are based either on model calculations in the weak ISS regime assuming that the SFs have all saturated, or the fitting formula of \citet{goodmannarayan06} with $v_{s} = 50\,{\rm kms}^{-1}$, $D_{\rm ISM} = 500 {\rm pc}$ and $\nu_{t} = 4.0$ GHz. The dotted horizontal line represents $R_{D} = 0.57$, which is the upper limit to $R_{D}$ fitted to all sources in the low redshift sample, at a confidence level of 99\%.} \label{igmconstraints}}
     \end{figure}

It has to be noted that Equation~\ref{EMSM} assumes a particular outer scale of Kolmogorov turbulence, and gives only the upper bounds of EM (see \citet{cordeslazio03}). Therefore, Equation~\ref{vthalpha} in fact gives only the upper limit to $\nu_t$, and initial calculations give $4 \lesssim \nu_t \lesssim 80$ GHz with a median of $\sim$ 12 GHz. We know from Figure~\ref{histoSFxSFc} that this is not the case, so a factor of $0.25$ is multiplied to $\nu_t$ to reduce its values to a range of $1 \lesssim \nu_t \lesssim 20$ GHz with a median of $\sim$ 3 GHz.
  
  \clearpage
 \begin{deluxetable}{c c c c c c c c c c}

\tabletypesize{\scriptsize}
\tablecaption{List of sources and their observed properties.\label{data}}
\tablehead{
\colhead{Source} & \colhead{$S_{4.9}\, {(\rm Jy)}$} & \colhead{$S_{8.4}\, {\rm (Jy)}$} & \colhead{$\alpha_{4.9}^{8.4}$} & \colhead{$D_{4.9}({\rm 4d)}$} & \colhead{$D_{8.4}({\rm 4d})$} & \colhead{$R_{D}$} & \colhead{ID} & \colhead{$z$} & \colhead{$I_{\alpha}\, {\rm (R)}$} \\

\colhead{(1)} & \colhead{(2)} & \colhead{(3)} & \colhead{(4)} & \colhead{(5)} & \colhead{(6)} & \colhead{(7)} & \colhead{(8)} & \colhead{(9)} & \colhead{(10)}
}

\startdata
J0009+1513 & 0.15 & 0.12 & -0.44 & 1.89e-4 $\pm$ 1.55e-4 & 1.44e-4 $\pm$ 1.09e-4 &        -        & fsrq & 2.2 &   0.7 \\ 
J0017+5312 & 0.59 & 0.64 &  0.14 & 4.68e-4 $\pm$ 1.57e-4 & 1.53e-4 $\pm$ 4.86e-5 & 0.33 $\pm$ 0.15 & fsrq & 2.6 &  13.3 \\ 
J0017+8135 & 1.36 & 1.26 & -0.13 & 1.89e-5 $\pm$ 1.52e-5 & 4.89e-6 $\pm$ 5.48e-9 &        -        & fsrq & 3.4 &   2.2 \\ 
J0056+1625 & 0.19 & 0.23 &  0.33 & 2.40e-3 $\pm$ 8.13e-4 & 1.91e-3 $\pm$ 7.00e-4 & 0.80 $\pm$ 0.40 & bllc & 0.2 &   0.8 \\ 
J0108+0135 & 1.53 & 2.06 &  0.56 & 6.49e-4 $\pm$ 1.87e-4 & 7.94e-5 $\pm$ 7.61e-5 &        -        & fsrq & 2.1 &   0.7 \\ 
J0122+0310 & 0.11 & 0.11 & -0.04 & 3.19e-4 $\pm$ 1.47e-4 & 3.70e-4 $\pm$ 1.47e-4 & 1.16 $\pm$ 0.70 & fsrq & 4.0 &   0.5 \\ 
J0122+2502 & 0.75 & 0.66 & -0.21 & 6.65e-5 $\pm$ 4.59e-4 & 8.03e-7 $\pm$ 2.57e-4 &        -        & fsrq & 2.0 &   0.9 \\ 
J0126+2559 & 0.81 & 0.66 & -0.39 & 6.85e-5 $\pm$ 4.63e-5 & 3.25e-6 $\pm$ 1.02e-4 &        -        & fsrq & 2.4 &   1.0 \\ 
J0135+2158 & 0.18 & 0.14 & -0.37 & 1.02e-3 $\pm$ 3.74e-4 & 2.82e-4 $\pm$ 1.78e-4 & 0.28 $\pm$ 0.20 & fsrq & 3.4 &   0.9 \\ 
J0154+4743 & 0.50 & 0.60 &  0.35 & 7.53e-4 $\pm$ 1.58e-4 & 9.47e-4 $\pm$ 2.09e-4 & 1.26 $\pm$ 0.38 & fsrq & 1.0 &   8.6 \\ 
\enddata

\tablecomments{This table is published in its entirety in the electronic edition of the Astrophysical Journal. A portion is shown here for guidance regarding its form and content. (1) J2000.0 IAU name; (2) 4.9 GHz mean flux density; (3) 8.4 GHz mean flux density, may slightly differ from values published in Paper I following the removal of data points on days in which the 4.9 GHz subarray encountered data losses, to avoid biases; (4) Spectral index, may differ from values published in Paper I; (5) 4.9 GHz SF at time-lag of 4 days, the errors are 95\% confidence bounds in fitting the model; (6) 8.4 GHz SF at time-lag of 4 days, may differ from values published in Paper I, the errors are 95\% confidence bounds in fitting the model; (7) Ratio of $D_{8.4}({\rm 4d)}$ to $D_{4.9}({\rm 4d)}$, given only for $\geq 3 \sigma$ variable sources; (8) Optical identification, flat spectrum radio-loud quasar (fsrq), BL Lac object (bllc), Seyfert 1 galaxy (syf1), narrow-line radio galaxy (nlrg) or no ID available (null), obtained from the NASA Extragalactic Database (NED), SIMBAD database and Pursimo et al. (submitted); (9) redshift, obtained from the NASA Extragalactic Database (NED), SIMBAD database and Pursimo et al. (submitted); (10) line-of-sight H$\alpha$ intensity, obtained from \citet{haffneretal03}}. 

\end{deluxetable}

 \begin{deluxetable}{c c c c}

\tabletypesize{\footnotesize}
\tablewidth{0pt}
\tablecaption{Comparison of fitted $R_{D}$ values at low and high redshift. \label{variousRDz}}
\tablehead{
\colhead{AGN Class} & \colhead{Mean Flux Density} & \colhead{$z < 2$} & \colhead{$z > 2$}
}

\startdata
Type 1 				& $S_{4.9} < 0.3$ Jy 	& 0.39 $\pm$ 0.20 (27)	& 0.70 $\pm$ 0.46 (13) \\
Type 1				& $S_{4.9} \geq 0.3$ Jy	& 0.64 $\pm$ 0.51 (10)	& 1.25 $\pm$ 0.62 (11) \\
Type 0 \& Type 1    & $S_{4.9} < 0.3$ Jy 	& 0.38 $\pm$ 0.19 (29)	& 0.70 $\pm$ 0.46 (13) \\
Type 0 \& Type 1 	& $S_{4.9} \geq 0.3$ Jy	& 0.83 $\pm$ 0.45 (19)	& 1.25 $\pm$ 0.62 (11) \\
\enddata

\tablecomments{The numbers in brackets in the third and fourth columns indicate the number of sources in each category.}

\end{deluxetable}

 \begin{deluxetable}{l c c}

\tabletypesize{\footnotesize}
\tablewidth{0pt}
\tablecaption{Input parameters for ISS model in Section~\ref{expansion}. \label{inputparam1}}
\tablehead{
\colhead{Parameter} & \colhead{Symbol} & \colhead{Value}
}

\startdata
Scattering screen distance from Earth  & $D_{\rm ISM}$  				&		500 pc     \\
Scattering screen velocity & $v_{s}$   						&		50 ${\rm kms^{-1}}$	\\
Transition frequency between weak and strong ISS & $\nu_{t}$ &		4.0 GHz			\\
Source intrinsic brightness temperature & $T_{b,int}$		&		$10^{11}$ K		\\
Source compact fraction & $f_{c}$							&		0.5				\\
4.9 GHz and 8.4 GHz mean flux density (strong sources) & $S_{\nu}$ 	&		1.00 Jy\\
4.9 GHz and 8.4 GHz mean flux density (weak sources) & $S_{\nu}$	&		0.15 Jy\\

\enddata

\end{deluxetable}

\end{document}